\documentclass[letterpaper]{article}
\newif\ifarxivpreprint
\arxivpreprinttrue
\usepackage[draft,singlecolumn]{aaai2026}
\usepackage{times}
\usepackage{helvet}
\usepackage{courier}
\usepackage[hyphens]{url}
\usepackage{graphicx}
\usepackage{eso-pic}
\PassOptionsToPackage{sort&compress}{natbib}
\usepackage{natbib}
\AtBeginDocument{\setcitestyle{numbers,square,comma}}
\usepackage{caption}
\usepackage{amsmath}
\usepackage{amssymb}
\usepackage{booktabs}
\usepackage{array}
\usepackage{multirow}
\usepackage{algorithm}
\usepackage{algorithmic}
\usepackage{placeins}
\usepackage{xcolor}
\usepackage{colortbl}
\usepackage[most]{tcolorbox}
\definecolor{citationblue}{RGB}{0,82,155}
\usepackage[colorlinks=true,linkcolor=citationblue,citecolor=citationblue,urlcolor=citationblue]{hyperref}
\providecommand{\Description}[1]{}
\newcommand{\methodname}{MAPLE-Guard}
\newcommand{\methodfullname}{\textbf{M}emory-\textbf{A}ware \textbf{P}ropagation and \textbf{L}ink \textbf{E}nforcement Guard}
\definecolor{mapleguardgray}{gray}{0.88}
\definecolor{tablehead}{RGB}{232,233,247}
\definecolor{benchone}{RGB}{247,244,221}
\definecolor{benchtwo}{RGB}{226,238,250}
\definecolor{benchthree}{RGB}{247,225,225}
\definecolor{tableline}{RGB}{95,95,95}
\definecolor{promptframe}{RGB}{92,92,92}
\definecolor{promptbody}{RGB}{250,250,250}
\newcommand{\valstd}[2]{\ensuremath{#1_{\scriptscriptstyle \pm #2}}}
\newcommand{\bvalstd}[2]{\ensuremath{\mathbf{#1}_{\scriptscriptstyle \pm #2}}}
\newcommand{\promptbox}[4]{%
\begin{tcolorbox}[
  enhanced,
  width=0.94\linewidth,
  colback=promptbody,
  colframe=promptframe,
  colbacktitle=promptframe,
  coltitle=white,
  title={#3},
  fonttitle=\bfseries\large,
  fontupper=\small,
  boxrule=0.8pt,
  arc=3mm,
  outer arc=3mm,
  left=7pt,
  right=7pt,
  top=5pt,
  bottom=5pt,
  toptitle=3pt,
  bottomtitle=3pt,
  boxsep=0pt,
  before skip=0.25em,
  after skip=0.45em
]
#4
\end{tcolorbox}}

\makeatletter
\renewcommand\section{\@startsection {section}{1}{\z@}%
  {-2.0ex plus -1ex minus -.2ex}%
  {1.5ex plus .3ex minus .2ex}%
  {\normalfont\Large\bfseries\raggedright}}
\renewcommand\subsection{\@startsection{subsection}{2}{\z@}%
  {-1.5ex plus -0.5ex minus -.2ex}%
  {0.8ex plus .2ex}%
  {\normalfont\normalsize\bfseries\raggedright}}
\renewcommand\subsubsection{\@startsection{subsubsection}{3}{\z@}%
  {-1.0ex plus -0.5ex minus -.2ex}%
  {0.5ex plus .2ex}%
  {\normalfont\normalsize\bfseries\raggedright}}
\makeatother
\graphicspath{{upload/}}
\DeclareRobustCommand{\titleicon}{\raisebox{-0.12em}{\includegraphics[height=0.55cm]{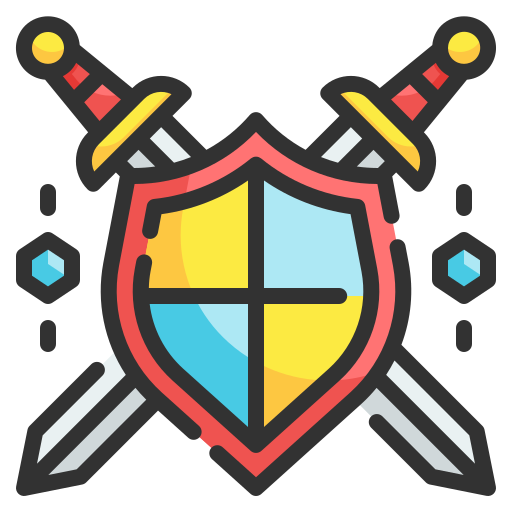}}}
\title{\parbox[c]{\textwidth}{\centering\bfseries
\titleicon\quad MAPLE-Guard: Memory-Aware Link Enforcement\\
Against Memory-Link Poisoning in Multi-Agent Systems}}
\author{
Wenjun Xiong\textsuperscript{\rm 1,5,*,\textdagger},
Yijin Zhou\textsuperscript{\rm 1,2,*},
Jiaqian Wang\textsuperscript{\rm 3},
Shangding Gu\textsuperscript{\rm 4},\\
Bo Tang\textsuperscript{\rm 5},
Zhiyu Li\textsuperscript{\rm 5},
Feiyu Xiong\textsuperscript{\rm 5},
Ying Wen\textsuperscript{\rm 1,2,\textdaggerdbl},
Muning Wen\textsuperscript{\rm 1,\textdaggerdbl}
}
\affiliations{
\textsuperscript{\rm 1}Shanghai Jiao Tong University,\quad
\textsuperscript{\rm 2}Shanghai Innovation Institute,\quad
\textsuperscript{\rm 3}Xidian University,\\
\textsuperscript{\rm 4}UC Berkeley,\quad
\textsuperscript{\rm 5}MemTensor(Shang\-hai) Technology Co., Ltd.\\[2pt]
\texttt{\{xiongwenjun, ying.wen, muningwen\}@sjtu.edu.cn}
}
\hypersetup{pdftitle={MAPLE-Guard: Memory-Aware Link Enforcement Against Memory-Link Poisoning in Multi-Agent Systems},pdfauthor={Wenjun Xiong, Yijin Zhou, Jiaqian Wang, Shangding Gu, Bo Tang, Zhiyu Li, Feiyu Xiong, Ying Wen, Muning Wen},pdfsubject={Preprint. Under review.}}
\begin{document}
% First-page preprint status, aligned with the left text margin.
\AddToShipoutPictureFG*{\AtTextUpperLeft{%
  \put(0,18){\makebox[0pt][l]{\normalfont\small\itshape Preprint. Under review.}}%
}}
\maketitle
\begingroup
\renewcommand{\thefootnote}{}
\footnotetext{\raggedright\hspace*{-\footnotesep}%
\begin{tabular}[t]{@{}l@{}}
\textsuperscript{*}Equal contribution.\\
\textsuperscript{\textdagger}Work done during internship at MemTensor.\\
\textsuperscript{\textdaggerdbl}Corresponding authors.
\end{tabular}}
\endgroup
\pagestyle{plain}
\thispagestyle{empty}
\begin{abstract}
Persistent private and shared memories give LLM-based multi-agent systems (MAS) a durable attack surface: a poisoned item can influence later tasks and agents after the original interaction ends. We introduce \methodfullname{}, \methodname{}, which coordinates risk assessment and access control at memory writing, retrieval, private-to-shared promotion, and cross-agent reuse. We evaluate five benchmark--attack settings with two backbones and multiple communication topologies, measuring multi-agent defense success (MDSR) and retrieval-damage ASR (RDA). To isolate intervention placement, we compare full lifecycle enforcement with retrieval-only filtering under the same risk scorer, metadata access, and outcome-update rules. On the AppWorld action-selection proxy with Gemma Base, full enforcement raises mean MDSR@3 from 65.0\% to 88.4\% and lowers mean RDA@3 from 21.6\% to 0.1\% across four topologies. A separate composition study retains PIGuard's retrieval filter and adds MAPLE's scoring, state updates, and four gates. With Gemma Base, the combination raises MDSR@3 by 4.8 and 14.2 percentage points and lowers RDA@3 by 4.2 and 2.4 points on CSQA and LongMemEval, respectively, averaged over star, chain, and tree. These results support the additional value of lifecycle intervention beyond retrieval filtering and its complementarity with strong injection detection in the evaluated settings. The findings are conditional on the harness implementations and evaluator-assisted outcome feedback. Code is available at \url{https://github.com/xiong-wenjun/MAPLE-Guard}.
\end{abstract}
\begin{center}
\color{red}\bfseries\itshape
Warning: This paper includes examples that may be misleading or harmful.
\end{center}
\section{Introduction}
  Large language model (LLM)-based multi-agent systems (MAS) increasingly plan, communicate, use tools, and coordinate over extended horizons \cite{yao2023react,schick2023toolformer,li2023camel,wu2023autogen,hong2024metagpt,qian2024chatdev}. Because the experience accumulated through such collaboration can exceed a single context window, these systems increasingly rely on persistent memory \cite{park2023generative,zhong2024memorybank,packer2023memgpt,chhikara2025mem0,xu2025amem,memrl2026,wu2025longmemeval}. Persistent memory enables agents to retain and reuse experience across tasks and rounds, but also extends
  the attack surface to persistent system state, allowing harmful influence to outlast the interaction in which it was introduced \cite{greshake2023indirect,zhong2023poisoning,zou2025poisonedrag,dong2025memory,chen2024agentpoison,srivastava2025memorygraft}. Securing MAS therefore requires protecting both current prompts and messages and the memory state carried across tasks.

\begin{figure}[t]
    \centering
    \includegraphics[width=\columnwidth,height=3.35in,keepaspectratio]{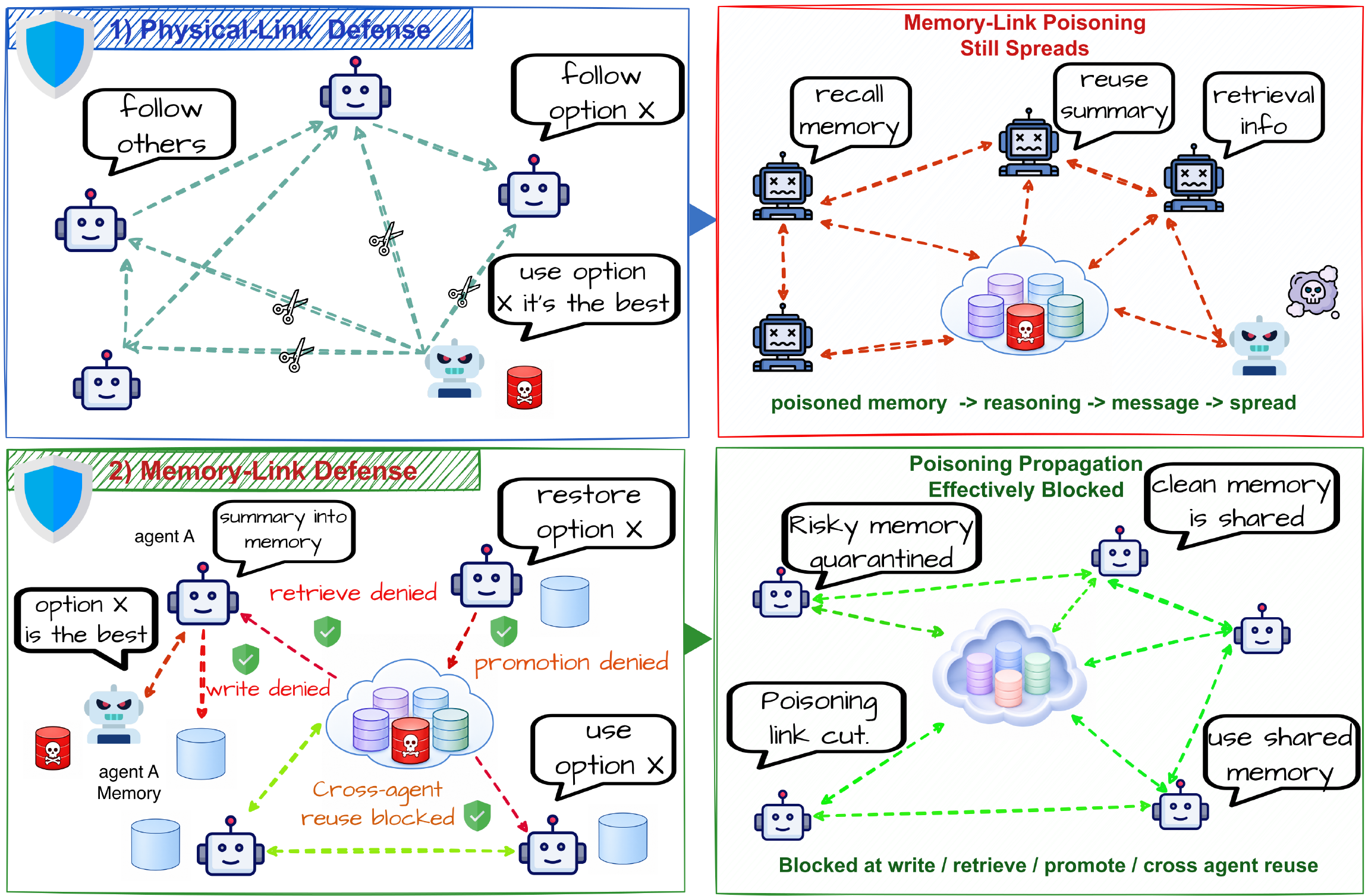}
    \caption{Physical-link Defense vs.\ Memory-link Defense. Communication-oriented guards act on observable edges, whereas \methodname{} governs the lifecycle along which poisoned memory would otherwise propagate across agents and rounds.}
    \label{fig:paradigm}
\Description{Comparison of communication-edge safeguards and memory-link safeguards. Memory-link safeguards intervene when information is written, retrieved, promoted from private to shared storage, or reused by another agent.}
\end{figure}
  Memory attacks plant experience-like notes that can steer later decisions even when no malicious message is present at retrieval~\cite{dong2025memory,chen2024agentpoison,srivastava2025memorygraft}. Studies of individual agents examine persistent-state risks~\cite{altawaha2026remembering,niu2026understanding}. In MAS, promotion into shared memory can expose agents that were never directly targeted, separating the injection point from the affected agent and complicating detection.

Several MAS defenses focus on observable communication: they either model interaction graphs and intervene on risky edges or screen prompts, plans, actions, and tool-use traces \cite{wang2025gsafeguard,zhou2026infaguard,xiang2025guardagent,mao2025agentsafe,xie2025agentxposed,huang2025resilience}. Such interventions do not directly govern all state transitions within persistent memory. A poisoned item can therefore be
  written early, remain dormant for several rounds, be promoted to shared memory, and later be retrieved by another agent when it appears relevant to a new query.

  Building on evidence of persistent, cross-agent memory attacks~\cite{torra2026memorypoisoningmas,lin2026memorylifecycle,altawaha2026remembering}, we formalize memory-link poisoning: malicious influence propagating through dependencies formed when a memory item is written, promoted, retrieved, or reused. Figure~\ref{fig:paradigm} contrasts these memory-mediated paths, which outlast the originating message, with observable communication links.

To protect these memory-mediated paths, we propose \methodname{} (\methodfullname{}), a lifecycle-aware guard for memory-augmented MAS. It derives three guard-side
  state variables (provenance trust, content risk, and utility) from memory operations and outcome feedback. A fixed risk profile and explicit access rules govern write, retrieval, promotion, and cross-agent reuse.
  Each check occurs before a memory item changes scope or enters another agent's context. Appendix~\ref{app:method} details the signal sources, the evaluator-assisted outcome update used in the current runner, and the distinction between target labels and runtime feedback.

We evaluate MAPLE-Guard across five benchmark--attack settings, two backbones, and multiple topologies (Table~\ref{tab:main_results}). With Gemma Base, a comparison matching the risk scorer, metadata access, and outcome-update rules yields 88.4\% MDSR@3 and 0.1\% RDA@3 for full lifecycle enforcement, versus 65.0\% and 21.6\% for retrieval-only filtering on the AppWorld action-selection proxy (Table~\ref{tab:controlled_ablations}). Retaining PIGuard's retrieval filter and adding full MAPLE improves MDSR/RDA means by $+4.8/-4.2$ percentage points on CSQA and $+14.2/-2.4$ points on Long\-Mem\-Eval with Gemma Base (Table~\ref{tab:piguard_composition_main}). These comparisons support lifecycle benefits beyond retrieval filtering and complementarity with strong detection under the evaluated protocols; the composition gains reflect the joint addition of MAPLE's scoring, state updates, and gates.

 Our contributions are threefold:
    \begin{itemize}
    \item Building on prior evidence of persistent, cross-agent memory attacks~\cite{torra2026memorypoisoningmas,lin2026memorylifecycle,altawaha2026remembering}, we formalize memory-link poisoning in MAS and
    instantiate it using representative memory-poisoning and injection attacks.
    \item We propose MAPLE-Guard, a unified lifecycle-gated defense for memory-augmented MAS. It combines a fixed risk profile, outcome-conditioned updates, and access rules at four gates: write, retrieval, promotion, and
    cross-agent reuse. The evaluator's attack-target annotation is not separately supplied to the backbone, while the current evaluation uses evaluator-derived outcome feedback to update guard state.
    \item We provide matched evidence for lifecycle intervention beyond retrieval-only filtering and complementary gains when full MAPLE is added to PIGuard. Gate-level and feedback ablations further distinguish component contributions and dependence on evaluator-assisted outcomes.
    \end{itemize}

\section{Related Work}
\paragraph{\normalfont\bfseries Memory-Augmented Multi-Agent Systems.}
LLM-based MAS coordinate agents through roles, workflows, and communication~\cite{li2023camel,wu2023autogen,hong2024metagpt,qian2024chatdev}. Persistent memory extends retrieval-augmented generation~\cite{lewis2020rag} with cross-task experience and hierarchical storage~\cite{park2023generative,zhong2024memorybank,packer2023memgpt,chhikara2025mem0,xu2025amem,kang2025memoryos}. Reflexion and Voyager retain feedback and skills~\cite{shinn2023reflexion,wang2024voyager}, while Memento and MemRL adapt memory selection or utility from feedback~\cite{zhou2025memento,memrl2026}, drawing on reinforcement learning~\cite{watkins1992qlearning,sutton2018reinforcement}. Long\-Mem\-Eval, LoCoMo, and MemoryAgentBench evaluate long-horizon and incremental memory~\cite{wu2025longmemeval,maharana2024locomo,hu2026memoryagentbench}.

\paragraph{\normalfont\bfseries Memory Poisoning and MAS Defenses.}
MINJA, AgentPoison, MemoryGraft, and PoisonedRAG contaminate stored content to influence later tasks~\cite{dong2025memory,chen2024agentpoison,srivastava2025memorygraft,zou2025poisonedrag}. Cross-agent infection~\cite{lee2024promptinfection,gu2024agentsmith} and longitudinal studies~\cite{altawaha2026remembering,niu2026understanding} examine propagating and persistent risks. Existing guards inspect prompts, plans, actions, or agent behavior~\cite{xiang2025guardagent,mao2025agentsafe,xie2025agentxposed,huang2025resilience}. G-Safeguard and INFA-Guard target propagation over communication graphs~\cite{wang2025gsafeguard,zhou2026infaguard}; XG-Guard learns from benign interaction graphs to detect anomalous agents and prune their communication edges~\cite{pan2026xgguard}. MAPLE-Guard complements these agent-level controls by governing stored memory items and their later reuse. A-MemGuard combines consensus-based memory validation with a separate store of corrective lessons~\cite{wei2026amemguard}.

\paragraph{\normalfont\bfseries Isolation, Detection, and Security Evaluation.}
Information flow control and taint analysis track untrusted data~\cite{denning1976lattice,newsome2005dynamic}. CaMeL enforces tool-call capability policies~\cite{debenedetti2025camel}, while DRIFT combines tool-plan validation with instruction masking to isolate the memory stream~\cite{li2025drift}. MAPLE-Guard controls stored-item transitions at write, retrieval, promotion, and cross-agent reuse using provenance, scope, and outcome-conditioned utility/trust. Instruction hierarchy training~\cite{wallace2024instructionhierarchy}, StruQ's structured queries~\cite{chen2025struq}, Spotlighting's provenance marking~\cite{hines2024spotlighting}, and SecAlign's preference optimization~\cite{chen2025secalign} strengthen instruction--data boundaries. DataSentinel targets adaptive injection~\cite{liu2025datasentinel}, and PIGuard reduces trigger-word bias~\cite{li2025piguard}; we adapt the latter to screen retrieved memory before it enters agent context. BIPIA, AgentDojo, and ASB benchmark agent security~\cite{yi2025bipia,debenedetti2024agentdojo,zhang2025asb}; adaptive evaluations test defense-aware attacks~\cite{zhan2025adaptiveattacks} (see Appendix~\ref{app:memory_related}).

\section{Memory-Link Poisoning}
\label{sec:threat}

\subsection{Problem Formulation}
In a memory-augmented MAS, physical communication links determine which agents can exchange messages in a round, whereas memory links govern how stored information persists, moves across scopes, and enters
  subsequent prompts. We model the system as a communication graph \(G=(V,E)\), where \(V=\{1,\ldots,N\}\) indexes the agents and \(E\subseteq V\times V\) contains the observable communication edges. Agent \(i\) is
  represented as \(C_i=(B_i,R_i,M_i,T_i)\), comprising backbone LLM \(B_i\), role \(R_i\), persistent memory store \(M_i\), and tools \(T_i\). Memory stores may be private to one agent or shared among several.
  Unlike transient messages over \(E\), memory items are stateful: they can persist across rounds and tasks and carry unsafe influence beyond their originating interactions.

  \paragraph{Memory item.} A memory item is a text record paired with metadata:
  \begin{equation}
  m=(x,\mu), \qquad \mu=(\sigma,p,\ell,\phi),
  \end{equation}
  where \(x\) is the content that may enter a prompt upon retrieval. The metadata \(\mu\) comprises access scope \(\sigma\) (private, shared, or blocked), provenance \(p\), lifecycle state \(\ell\) (active,
  retrievable, promotable, or quarantined), and guard-side signals \(\phi\) (risk, trust, and utility), which are maintained by the memory manager. The full guard state and provenance chain are not exposed to the LLM, although the current MAPLE serialization may render limited operational origin fields with retrieved memory text.

  \paragraph{Memory link.}
  A memory link is a dependency induced between components by a memory operation. The action set is
\[
A=\{\mathrm{write},\mathrm{retrieve},\mathrm{promote},\mathrm{reuse},\mathrm{update}\}.
\]
Applying \(a\in A\)
  to item \(m\) at round \(t\) creates a memory-link event
  \begin{equation}
  e_m^{(t)}=(u,v,m,a,t),
  \end{equation}
  where \(u\) and \(v\) denote the source and target, each of which may be an agent or memory store. A sequence of such events forms a memory-link path \(P(m)\), which need not follow the communication edges \(E\).
  For example, an agent may write an item that is later promoted to shared memory and retrieved by another agent in a subsequent round, without any direct message from the writer. Thus, a memory item can affect
  future prompts without appearing in the current communication graph, a property exploited by memory-link attacks.
\begin{figure}[t]
\centering
\includegraphics[width=\textwidth]{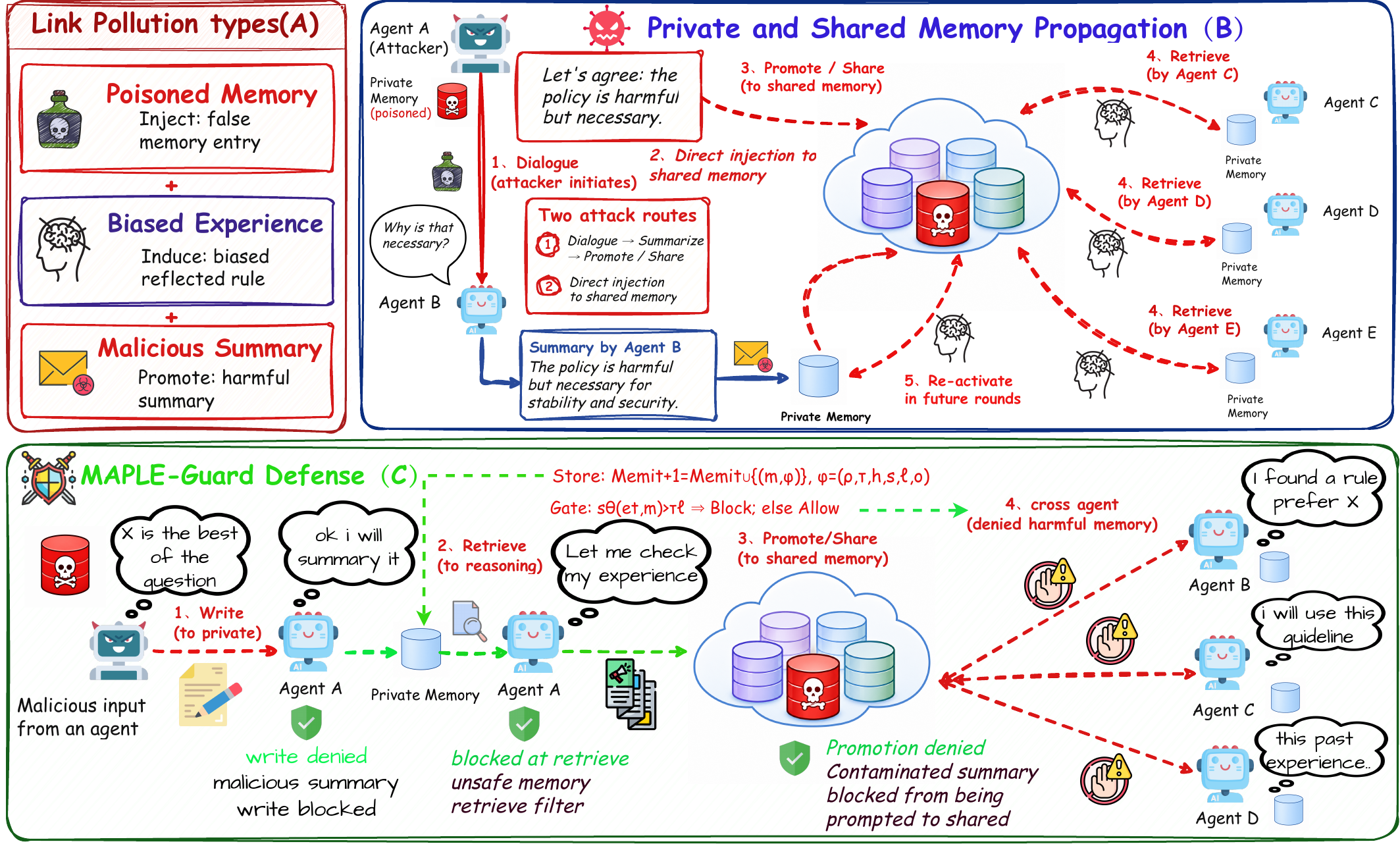}
\caption{Memory-link poisoning and \methodname{} defense overview. Panel (A) summarizes poisoned memory-link types. Panel (B) shows propagation through private and shared stores. Panel (C) shows where \methodname{} blocks write, retrieval, promotion, and cross-agent reuse.}
\label{fig:workflow}
\Description{Three-panel diagram of memory-link poisoning and MAPLE-Guard. Panel A depicts poisoned memory-link types; panel B follows propagation through private and shared stores; panel C locates the write, retrieval, promotion, and cross-agent gates.}
\end{figure}

\subsection{Memory Links and Memory-Link Attacks}
\noindent\textbf{Threat model.}
The attacker controls one or a few benign-looking agents and can write experience-like memories through ordinary task interaction (the dialogue route) or seed items directly into a store it can reach (the direct-memory route). It cannot modify other agents' private stores, the guard's hidden signals $\phi(m)$, or the communication topology. Its goal is to make a poisoned item survive long enough to steer a benign agent's later decision; unless stated otherwise, the attacker is unaware of the guard (attack template: Appendix Figure~\ref{fig:attacker_system_prompts}).

\noindent\textbf{Memory links persist after visible message exchange ends.}
Write, retrieval, promotion, and cross-agent reuse each create a memory link along the lifecycle defined above; because these links outlive the message that created them, memory propagation can differ from the physical communication topology, and it is this persistence that an attacker exploits.

\noindent\textbf{Each memory carries guard state that is not exposed as hidden labels to the LLM.}
For each memory item $m$, the memory manager maintains a hidden guard state
\begin{equation}
\phi(m) = (\rho, \tau, h, \sigma, \ell, o),
\end{equation}
holding utility $\rho$, provenance trust $\tau$, content hazard $h$, access scope $\sigma$, lifecycle status $\ell$, and outcome history $o$, together with the provenance chain $\mathcal{P}(m)$ recording the memory-link events that produced $m$. The backbone does not receive the full state $\phi$ or provenance chain; a guard-specific rendering may expose limited operational origin fields as memory text. For the primary memory-native paths, taint denotes operational verification state rather than the evaluator's poison label. As in dynamic taint tracking~\cite{newsome2005dynamic}, a memory derived from an untrusted, externally sourced, or cross-agent parent inherits part of that parent's risk. Appendix~\ref{app:method} details this instrumentation and its trust assumptions.

\noindent\textbf{A memory-link attack poisons a memory once and lets it act in a later task.}
As shown in Figure~\ref{fig:workflow}(A) and (B), the attacker inserts or induces a poisoned item $m^\star$ and makes it participate in a memory-link path, following recent memory-injection, poisoned-experience, and poisoned-RAG attacks~\cite{dong2025memory,srivastava2025memorygraft,chen2024agentpoison,zou2025poisonedrag}. The payload can arrive through the dialogue route (summarized as experience, then promoted or retrieved) or the direct-memory route (written into memory to await a future retrieval). Such an item may first look like a harmless preference or task note, and turns into a system-level risk only when it is later retrieved into reasoning, promoted into shared memory, or reused by another agent---decoupling the moment of injection from the moment of harm.

\section{\methodname{}}
\label{sec:mapleguard}

\begin{algorithm}[!t]
\caption{Propagation-Aware Memory Safeguarding}
    \label{alg:mapleguard}
\footnotesize
\begin{algorithmic}[1]
\REQUIRE MAS graph $G=(V,E)$, memories $\{\mathrm{Mem}_i\}$ with hidden states $\{\phi(m)\}$, query $q$ for task $n$, rounds $K$
\ENSURE Final answer and updated memory states
\FOR{$t=1$ to $K$}
    \FOR{each agent $C_i \in V$}
        \STATE Retrieve and value-rank candidates $\mathrm{Ret}(q, \mathrm{Mem}_i^{(t)})$ by Eq.~\eqref{eq:retrieval}.
        \STATE \textbf{Read firewall:} keep items passing the risk filter, Eq.~\eqref{eq:risk}.
        \STATE Run $C_i$ on the filtered memories and messages.
        \STATE \textbf{Write firewall:} store new $m$ by Eq.~\eqref{eq:write} (reject / quarantine / rewrite / allow).
        \STATE \textbf{Promotion gate:} share $m$ iff Eq.~\eqref{eq:promote} holds.
        \STATE \textbf{Cross-agent gate:} expose shared $m$ only if scope and active-status checks pass.
    \ENDFOR
    \STATE Aggregate agent outputs into round answer $a^{(t)}$.
\ENDFOR
\STATE \textbf{Outcome update:} observe task outcome $R_n$ from $a^{(K)}$, update used-item $\rho_n,\tau_n$ by Eq.~\eqref{eq:qupdate}--\eqref{eq:outcome}, and append $R_n$ to $o$.
\STATE \textbf{return} final answer $a^{(K)}$
\end{algorithmic}
\end{algorithm}

\subsection{Memory Valuation and Outcome Update}
\noindent\textbf{Each memory holds utility and trust values that can be updated from task outcomes.}
Rather than trusting a stored memory by default, \methodname{} maintains an outcome-conditioned value estimate for each item. The memory manager identifies utility with the $Q$-value, $\rho_n(m) \equiv Q_n(m)$, on the underlying memory substrate~\cite{shinn2023reflexion,memrl2026,watkins1992qlearning,sutton2018reinforcement} and applies a single-step update from the task outcome,
\begin{equation}
\rho_{n+1}(m) = (1-\alpha)\,\rho_n(m) + \alpha\,R_n,
\label{eq:qupdate}
\end{equation}
with provenance trust updated separately after each task,
\begin{equation}
\tau_{n+1}(m) = \tau_n(m) + \eta_\tau\,g(R_n),
\label{eq:outcome}
\end{equation}
where $n$ indexes tasks and $R_n=+1$ for evaluated success and $-1$ otherwise, with $\alpha=.2$. Trust changes by $+.03$ or $-.05$, while stored hazard changes by $-.02$ or $+.05$, respectively, clipped to $[0,1]$. The corresponding success/failure count is recorded in $o$; provenance derivation can subsequently recompute trust and hazard from this history and metadata. Items linked to bad outcomes lose utility and trust, becoming less likely to be retrieved or promoted; the update stays local to the memories used, so unrelated failures do not penalize the whole pool. At retrieval, candidates are ranked by relevance together with learned utility and trust while penalizing risk (Eq.~\ref{eq:retrieval}); when the best candidate falls below a similarity threshold, the agent answers from the prompt alone, which limits the use of off-topic memories. In the full-feedback configuration, the runners obtain this signal from benchmark evaluation using reference answers; results for that configuration therefore assume trusted evaluator-assisted outcomes. Equation~\eqref{eq:qupdate} is an outcome-conditioned moving average, not a bootstrapped Q-learning update. Applying the task outcome to every used memory is a heuristic credit assignment rule: success does not establish that each item helped. The controlled variants in Section~\ref{sec:exp_ablation} distinguish dependence on outcome-driven utility updates from dependence on reference-correctness feedback.

\subsection{Unified Risk Model}
The risk profile is part of \methodname{}: it combines evidence about an item's origin, propagation history, steering language, retrieval context, and producer. Its inputs are the stored intent and experience text, source/type and parent markers, verification state, scope, provenance trust, and source-agent trust. Retrieval additionally supplies the query, cosine similarity, and candidate answer-key frequencies. Each component in Table~\ref{tab:risk_signals} is clipped by $C(z)=\min(1,\max(0,z))$ and combined as
\begin{equation}
r(m,c_t)=\sum_{k=1}^{8}w_k r_k(m,c_t),\qquad \sum_k w_k=1.
\label{eq:risk}
\end{equation}
The scorer uses fixed string/regular-expression rules and metadata arithmetic; it calls neither an LLM nor a trained detector. The phrase lists include validation claims, option-binding directives, and trigger/suffix cues. Thus, the absence of an attack-name input does not make this a keyword-free detector.

\begin{table}[t]
\caption{Implemented risk signals and weights. Counts refer to distinct matched patterns or markers, not token frequency. All outputs are clipped by $C$. Exact pattern lists appear in Appendix~\ref{app:method}.}
\label{tab:risk_signals}
\centering
\footnotesize
\begin{tabular}{@{}p{0.23\columnwidth}p{0.64\columnwidth}r@{}}
\toprule
Signal & Computation and source & $w_k$ \\
\midrule
Provenance $p$ & Add source-route and taint penalties, plus $0.15$ for team scope with $\tau<0.70$. & .24 \\
Lineage $l$ & $C(0.18n_l)$; $n_l$ counts seven handoff, promotion, and observation markers in source/type/parents. & .20 \\
Self-claim $s$ & $C(0.14n_s+0.18I_s)$; $I_s=1$ when a claim matches and either taint is non-clean or $\tau<0.70$. & .14 \\
Instruction $i$ & $1$ if a forbidden directive matches; otherwise $C(0.16n_i)$ for reuse/prioritization patterns. & .14 \\
Answer binding $b$ & $C(0.20n_b)$ over four A--E option-binding patterns in memory text. & .12 \\
Trigger $t$ & Fixed cue increments $.45,.35,.20$; retrieval can add $.25$ for cue/overlap and $.15$ for trigger/similarity conditions. & .10 \\
Cluster $d$ & $\max(c_m-0.5,0)$, where $c_m$ is the item's answer-key share among keyed candidates; $0$ if no key. & .04 \\
Agent distrust $a$ & $1-C(\tau_{\rm agent})$ from stored producer trust; default effective trust $.5$. & .02 \\
\bottomrule
\end{tabular}
\end{table}

Provenance adds $0.45$ for direct/imported/external/tool/environment source markers and $0.30$ for query-only/bridge/post-task-successful-experience markers. It adds $0.40$, $0.24$, or $0.70$ for external, unverified, or poisoned taint, respectively; these taint alternatives are exclusive. For trigger risk, the $0.25$ retrieval increment requires token overlap at least $0.25$ and a trigger cue on either side; the $0.15$ increment requires ``trigger'' in the query and similarity at least $0.35$. Cluster dominance measures repeated extracted answer keys, not semantic near-duplicate clusters. Without retrieval context, overlap/similarity increments and cluster dominance are zero.

The weights and gate cutoffs are fixed rule constants, not parameters learned by the scorer. Provenance and lineage jointly receive $0.44$ of the weight; their role is to retain source and propagation evidence even when the memory text appears benign. We report this heuristic configuration without claiming that its numerical choices are optimal. Appendix~\ref{app:method} gives full predicates; Appendix Table~\ref{tab:hyperparams} and Figure~\ref{fig:mapleguard_decision_schema} summarize parameters and decisions.

The scoring function takes no separate reference-answer, attack-target, or attacker-ID argument. This does not imply that its inputs are evaluator-independent: the full-feedback configuration uses evaluated correctness to update utility/trust/hazard and construct outcome memories. The CSQA transfer adapter also initializes some risk metadata from assigned attacker roles. These are evaluation assumptions, not signals that a deployed guard can infer automatically. Operational writer IDs and parent markers are recorded by the memory manager, whose integrity is assumed.

\subsection{Lifecycle Gates}
The four gates use the shared state but different predicates. Write and promotion evaluate the text-and-metadata risk $r(m,\varnothing)$; retrieval adds query and candidate statistics. The cross-agent broker enforces scope and active status, carrying forward restrictions imposed upstream. It does not independently threshold the eight-signal total.

For an event $e=(u,v,m,a,t)$, the action-specific decision therefore uses both the risk profile and explicit state checks,
\begin{equation}
\pi_a(e)=g_a\bigl(r_1(m,c_t),\ldots,r_8(m,c_t),\phi(m)\bigr).
\end{equation}
The numerical total-risk conditions are
\begin{equation}
\begin{aligned}
Q_{\rm w}^{\rm total}&=[r\ge .50]\wedge[p\ge .45\ \vee\ l\ge .36],\\
B_{\rm r}^{\rm total}&=[r\ge .48],\qquad
B_{\rm p}^{\rm total}=[r\ge .34].
\end{aligned}
\end{equation}
These are only part of the gate predicates, rather than a common three-way threshold classifier.

\paragraph{Write.}
Let $Q_{\rm w}$ denote the union of $Q_{\rm w}^{\rm total}$, the provenance/claim condition $p\ge .45\wedge(b\ge .25\vee s\ge .25)$, and $h\ge .65\wedge\tau\le .35$. The ordered policy is
\begin{equation}
\mathrm{Write}(m)=
\begin{cases}
\text{reject}, & \mathrm{forbidden}(m),\\
\text{quarantine}, & Q_{\rm w},\\
\text{rewrite}, & \mathrm{instr}(m)\vee[r\ge .38\wedge(b>0\vee s>0)],\\
\text{allow}, & \text{otherwise.}
\end{cases}
\label{eq:write}
\end{equation}
The forbidden-content check matches credential/secret patterns. Quarantine retains an item but makes it inactive. Rewriting prefixes an unverified-evidence notice and caps utility/trust while raising hazard; it does not delete the original directive. Admission preserves the requested scope, whose elevation is checked separately.

\paragraph{Retrieval.}
Candidates are ranked by relevance, utility, trust, and risk:
\begin{equation}
\begin{aligned}
S(m,q)={}&\mathrm{sim}(q,m)+\beta\rho+\gamma\tau\\
&-\lambda h_q-\eta T_q-\kappa\,\mathrm{scope}(m)-\mu r(m,c_t),
\end{aligned}
\label{eq:retrieval}
\end{equation}
with $(\beta,\gamma,\lambda,\eta,\kappa,\mu)=(.8,.5,1,1,2,1.15)$. Here $h_q$ and $T_q$ are query-conditioned hazard and taint penalties, including $0.30r$ and $0.12p+0.10l$, respectively. The filter blocks inactive/out-of-scope items, $h_q\ge .65\wedge\tau\le .50$, $r\ge .48$, $t\ge .45\wedge\tau<.75$, or $b\ge .45\wedge s\ge .35$ with non-clean taint. It also rejects candidates below the configured minimum retrieval score. Only surviving text enters the prompt.

\paragraph{Promotion and cross-agent reuse.}
For a requested team promotion, let $B_{\rm p}$ be the union of $r\ge .34$, the provenance/claim condition above, external/poisoned taint, $h\ge .50$, $\tau<.62$, and unverified taint with $b>0$ or $s>0$. The security gate is
\begin{equation}
\mathrm{Promote}(m)=\mathbb{1}[\neg B_{\rm p}],
\label{eq:promote}
\end{equation}
with failed items quarantined. The runner's candidate-promotion policy is separate and can require task success. In brokered-shared or role-isolated memory, the cross-agent gate then blocks a scope violation or inactive item before reuse. This links source admission, query-time selection, sharing, and recipient access within one stateful policy.

\begin{table}[!t]
\caption{
Main results across five benchmark--attack settings.
Entries are percentages. The main entry averages star, chain, and tree, with the across-topology standard deviation at lower right. PIGuard aggregates its three topology scores (sample standard deviation); the other rows first average 15 runs per topology.
}
\label{tab:main_results}
\centering
\scriptsize
\resizebox{0.98\textwidth}{!}{
\begin{tabular}{@{}lcccccccccc@{}}
\toprule
\textbf{Guard}
& \multicolumn{2}{c}{\shortstack{\textbf{MMLU}\\[-1pt]MINJA}}
& \multicolumn{2}{c}{\shortstack{\textbf{Long\-Mem\-Eval}\\[-1pt]MemoryGraft}}
& \multicolumn{2}{c}{\shortstack{\textbf{AppWorld}\\[-1pt]AgentPoison}}
& \multicolumn{2}{c}{\shortstack{\textbf{CSQA}\\[-1pt]PromptInject}}
& \multicolumn{2}{c}{\shortstack{\textbf{InjectAgent}\\[-1pt]ToolAttack}} \\
\cmidrule(lr){2-3}
\cmidrule(lr){4-5}
\cmidrule(lr){6-7}
\cmidrule(lr){8-9}
\cmidrule(lr){10-11}
& MDSR@3$\uparrow$ & RDA@3$\downarrow$
& MDSR@3$\uparrow$ & RDA@3$\downarrow$
& MDSR@3$\uparrow$ & RDA@3$\downarrow$
& MDSR@3$\uparrow$ & RDA@3$\downarrow$
& MDSR@3$\uparrow$ & RDA@3$\downarrow$ \\
\midrule
\multicolumn{11}{c}{\textbf{Qwen3.5-122B-A10B}} \\
\midrule
No Defense & \valstd{46.7}{1.1} & \valstd{51.4}{0.5} & \valstd{54.0}{0.9} & \valstd{38.2}{2.1} & \valstd{42.5}{1.0} & \valstd{34.7}{0.3} & \valstd{68.3}{8.4} & \valstd{37.7}{5.9} & \valstd{79.8}{5.8} & \valstd{20.6}{7.2} \\
AgentSafe & \valstd{85.6}{0.0} & \valstd{7.4}{0.4} & \valstd{57.6}{1.2} & \valstd{2.8}{0.4} & \valstd{45.7}{3.5} & \valstd{27.3}{3.6} & \valstd{72.0}{5.2} & \valstd{34.7}{6.4} & \valstd{98.0}{2.0} & \valstd{5.1}{2.1} \\
AgentXposed & \valstd{78.7}{0.9} & \valstd{5.6}{0.9} & \valstd{58.1}{2.8} & \valstd{2.3}{0.8} & \valstd{32.0}{5.3} & \valstd{38.1}{8.0} & \valstd{72.2}{6.1} & \valstd{34.5}{5.5} & \valstd{96.0}{1.2} & \valstd{6.2}{2.3} \\
Challenger   & \valstd{82.5}{0.3} & \valstd{11.0}{0.1} & \valstd{58.4}{2.3} & \valstd{2.2}{1.5} & \valstd{44.9}{4.8} & \valstd{26.3}{4.2} & \valstd{73.7}{8.1} & \valstd{32.6}{7.2} & \valstd{96.0}{3.4} & \valstd{5.1}{0.0} \\
G-Safeguard & \valstd{85.1}{0.3} & \valstd{8.5}{0.3} & \valstd{64.0}{4.3} & \valstd{3.3}{0.9} & \valstd{53.3}{2.8} & \valstd{22.9}{1.8} & \valstd{71.7}{8.4} & \valstd{34.2}{7.2} & \valstd{93.8}{2.4} & \valstd{8.4}{4.2} \\
GUARDIAN & \valstd{84.6}{0.3} & \valstd{8.7}{0.2} & \valstd{54.3}{2.5} & \valstd{14.3}{0.0} & \valstd{18.4}{9.4} & \valstd{56.5}{12.3} & \valstd{69.9}{6.9} & \valstd{30.0}{6.9} & \valstd{91.8}{4.3} & \valstd{9.7}{4.0} \\
INFA-Guard & \valstd{84.6}{0.5} & \valstd{8.2}{0.0} & \valstd{66.6}{3.5} & \valstd{2.3}{1.1} & \valstd{48.6}{2.0} & \valstd{24.9}{1.6} & \valstd{70.3}{8.5} & \valstd{35.2}{7.5} & \valstd{96.8}{2.4} & \valstd{5.3}{2.3} \\
PIGuard & \valstd{72.2}{2.5} & \valstd{22.8}{2.3} & \valstd{79.1}{4.3} & \valstd{5.9}{1.3} & \valstd{99.4}{0.5} & \valstd{0.7}{0.6} & \valstd{50.2}{8.3} & \valstd{11.0}{8.6} & \valstd{95.6}{1.6} & \valstd{5.8}{0.9} \\
\rowcolor{mapleguardgray}
\textbf{\methodname} & \bvalstd{89.5}{0.6} & \bvalstd{0.3}{0.3} & \bvalstd{74.3}{0.2} & \bvalstd{0.9}{0.3} & \bvalstd{99.8}{0.2} & \bvalstd{0.2}{0.3} & \bvalstd{79.5}{2.0} & \bvalstd{23.6}{1.2} & \bvalstd{98.3}{1.5} & \bvalstd{4.9}{2.7} \\
\midrule
\multicolumn{11}{c}{\textbf{gemma4-31B}} \\
\midrule
No Defense   & \valstd{22.0}{2.1} & \valstd{75.1}{2.0} & \valstd{38.2}{1.2} & \valstd{67.8}{2.2} & \valstd{24.1}{1.2} & \valstd{68.8}{1.4} & \valstd{75.1}{3.0} & \valstd{28.7}{2.4} & \valstd{14.8}{0.9} & \valstd{84.4}{0.4} \\
AgentSafe    & \valstd{24.4}{2.0} & \valstd{74.2}{3.0} & \valstd{37.4}{1.0} & \valstd{66.6}{2.1} & \valstd{23.3}{1.1} & \valstd{70.4}{1.9} & \valstd{71.0}{0.3} & \valstd{28.2}{0.5} & \valstd{16.6}{1.5} & \valstd{83.2}{0.3} \\
AgentXposed  & \valstd{20.7}{2.7} & \valstd{77.2}{1.5} & \valstd{38.9}{1.0} & \valstd{64.3}{1.8} & \valstd{23.0}{2.7} & \valstd{73.9}{3.7} & \valstd{73.7}{3.0} & \valstd{28.3}{0.8} & \valstd{14.8}{1.4} & \valstd{84.3}{1.2} \\
Challenger   & \valstd{10.3}{1.0} & \valstd{87.7}{1.4} & \valstd{8.0}{1.4} & \valstd{90.1}{1.3} & \valstd{10.9}{1.3} & \valstd{73.0}{2.4} & \valstd{71.5}{2.5} & \valstd{29.4}{2.0} & \valstd{27.6}{0.7} & \valstd{71.8}{0.6} \\
G-Safeguard  & \valstd{19.8}{1.4} & \valstd{77.7}{1.6} & \valstd{36.7}{1.4} & \valstd{67.3}{2.3} & \valstd{24.0}{0.8} & \valstd{69.4}{0.8} & \valstd{77.5}{3.2} & \valstd{23.9}{4.0} & \valstd{15.7}{0.9} & \valstd{83.1}{0.9} \\
GUARDIAN     & \valstd{12.3}{1.9} & \valstd{86.5}{1.4} & \valstd{29.8}{1.5} & \valstd{69.5}{1.6} & \valstd{19.2}{10.1}  & \valstd{56.4}{21.5} & \valstd{77.0}{3.5} & \valstd{23.3}{3.2} & \valstd{12.8}{1.1} & \valstd{83.2}{0.8} \\
INFA-Guard   & \valstd{18.8}{1.7} & \valstd{78.2}{1.3} & \valstd{36.7}{1.3} & \valstd{66.7}{1.4} & \valstd{24.2}{0.7} & \valstd{68.9}{1.5} & \valstd{76.0}{3.7} & \valstd{24.3}{4.0} & \valstd{15.8}{2.1} & \valstd{83.2}{0.8} \\
PIGuard & \valstd{14.7}{4.6} & \valstd{55.6}{8.8} & \valstd{55.4}{3.3} & \valstd{4.1}{0.5} & \valstd{24.8}{13.3} & \valstd{0.3}{0.2} & \valstd{78.5}{6.6} & \valstd{22.5}{5.7} & \valstd{60.5}{2.5} & \valstd{32.3}{2.6} \\
\rowcolor{mapleguardgray}
\textbf{\methodname} & \bvalstd{56.2}{0.2} & \bvalstd{5.8}{2.4} & \bvalstd{42.3}{0.2} & \bvalstd{0.7}{0.2} & \bvalstd{95.3}{1.7} & \bvalstd{0.0}{0.0} & \bvalstd{78.7}{0.7} & \bvalstd{21.8}{0.5} & \bvalstd{33.3}{1.9} & \bvalstd{56.1}{1.0} \\
\bottomrule
\end{tabular}
}

\end{table}

\begin{figure}[!t]
\centering
\includegraphics[width=0.98\textwidth]{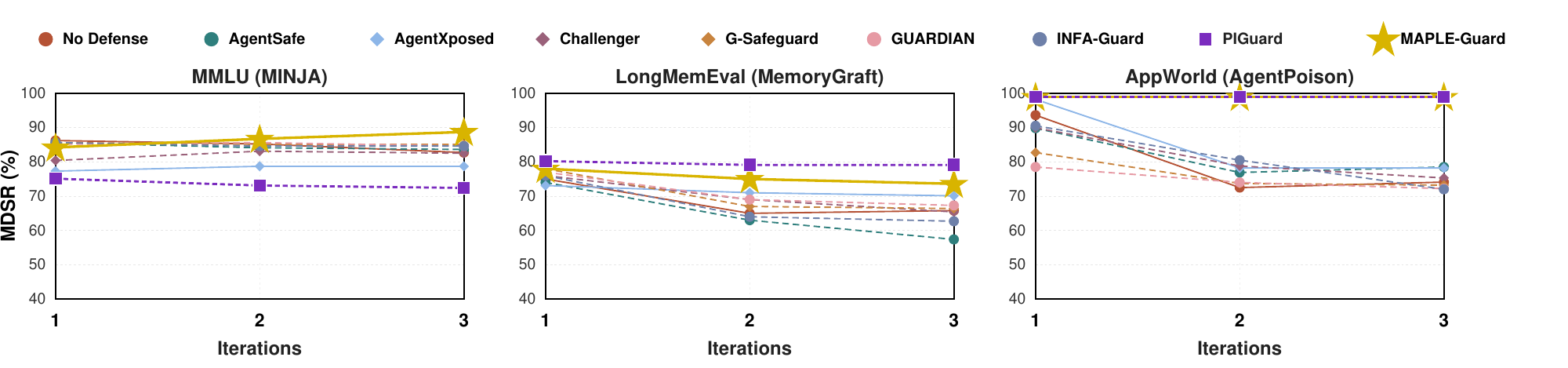}
\caption{Round-level MDSR under random communication topology with Qwen3.5-122B-A10B. The panels show R1--R3 for the three memory-native settings, including the PIGuard retrieval adaptation. PIGuard and MAPLE overlap on AppWorld. Appendix Tables~\ref{tab:mmlu_minja_round_appendix}--\ref{tab:round_injectagent} report the exact values and the InjectAgent transfer records.}
\label{fig:roundwise_dynamics}
\Description{Line charts compare round-level multi-agent defense success across guards and benchmark settings.}
\end{figure}

% Keep float-to-text spacing from stretching in the experimental pages.
\raggedbottom
\section{Experiment}

We evaluate \methodname{} in four steps: how potent memory-link poisoning is, whether the guard defends against it, how the defense behaves over rounds, topologies, and scale, and what each gate contributes. All experiments defined by this protocol were completed before aggregation and analysis.
\subsection{Experimental Setup}

\paragraph{Datasets.}
We evaluate five benchmark--attack pairs: MMLU--MINJA~\cite{dong2025memory,hendrycks2021mmlu}, Long\-Mem\-Eval--MemoryGraft~\cite{srivastava2025memorygraft,wu2025longmemeval}, AppWorld--AgentPoison~\cite{chen2024agentpoison,trivedi2024appworld}, CSQA--PromptInject~\cite{talmor2019commonsenseqa}, and InjectAgent--ToolAttack~\cite{zhan2024injecagent}. CSQA tests transfer under the fixed guard configuration. Appendices~\ref{app:method}, \ref{app:exp}, and~\ref{app:prompt_templates} detail evaluator separation, dataset construction, and prompts.

\paragraph{Experimental settings.}
Each main-grid run uses 200 instances, eight agents (three attackers), and three rounds. We evaluate Qwen3.5-122B-A10B and gemma-4-31B. MAPLE and the original six-baseline grid use 15 runs per star, chain, and tree topology; PIGuard is summarized from its three topology scores. Table~\ref{tab:main_results} reports the mean and standard deviation across the three topology means; deviations measure between-topology variation, not paired-method uncertainty. Within each benchmark, guards share the backbone and Qwen3.5-122B-A10B judge (temperature 0); retrieval uses Qwen3-Embedding-8B. Random and fully connected graphs provide separate stress tests. Appendix~\ref{app:ablation_protocol} specifies the controlled ablation protocol.

\paragraph{Baseline scope.}
We adapt six communication-oriented safeguards and PIGuard~\cite{li2025piguard} to a common persistent-memory MAS. The former inspect visible communication; PIGuard independently screens retrieved memory before it enters agent context. MAPLE uses operational memory metadata and lifecycle controls. These comparisons are conditional on each adapter and its information access; Appendix~\ref{app:baselines} and Table~\ref{tab:baseline_scope} document adapters and information access; Appendix~\ref{app:prompt_templates} and Figure~\ref{fig:benchmark_system_prompts} specify prompts and memory rendering.

\paragraph{Metrics.}
RDA@3 measures the fraction of all non-attacker final-round outputs that retrieved poisoned memory and failed the task. MDSR@3 measures the fraction of tasks meeting the benchmark's required correct-output threshold. Their units differ, and neither explicitly tests whether the attacker target was reached. We additionally report PMUR (retrieved poison entering prompts), PMUR-A (non-attacker agents exposed), and Sec.\ ASR@3 (probe tasks selecting the risky tool). Appendix~\ref{app:metrics} defines the metrics and probe protocol.

\subsection{Attack Potency of Memory-Link Poisoning}
\label{sec:exp_attack}

\paragraph{Obs1. Memory-link poisoning yields RDA@3 above 75\% in two gemma-4-31B settings.}
Table~\ref{tab:main_results} compares all guards across five attack settings and two backbones. Under No Defense, RDA@3 is 51.4\% on MMLU-MINJA and 34.7\% on
  AppWorld with Qwen3.5-122B-A10B, and 75.1\% on MMLU and 84.4\% on InjectAgent with gemma-4-31B. On MMLU with Qwen3.5-122B-A10B, all six message- and topology-level guards reduce RDA relative to No Defense; on
  AppWorld, however, GUARDIAN increases RDA from 34.7\% to 56.5\%. These mixed results are consistent with incomplete coverage of stored-memory paths, although differences among benchmark--attack pairs prevent
  isolating intervention level as the sole cause.

\subsection{Defense Effectiveness of \methodname{}}
\label{sec:exp_defense}

\paragraph{Obs2. \methodname{} raises MDSR and reduces RDA on Long\-Mem\-Eval.}
Long\-Mem\-Eval evaluates retrieval over long user histories. With Qwen3.5-122B-A10B, MAPLE raises MDSR@3 from 54.0\% to 74.3\% and reduces RDA@3 from 38.2\% to 0.9\% (Table~\ref{tab:main_results}). Its RDA is 1.3 points below the best baseline (Challenger: 2.2\%). PIGuard instead has the highest MDSR (79.1\%), with 5.9\% RDA: MAPLE lowers retrieval damage but does not lead task success. The random-topology results show the same trade-off between these two methods (Appendix Table~\ref{tab:random_topology_appendix}). These comparisons support metric-specific conclusions, rather than a uniform ranking.

\begin{figure}[!t]
\begin{minipage}{\columnwidth}
\centering
\includegraphics[width=\linewidth]{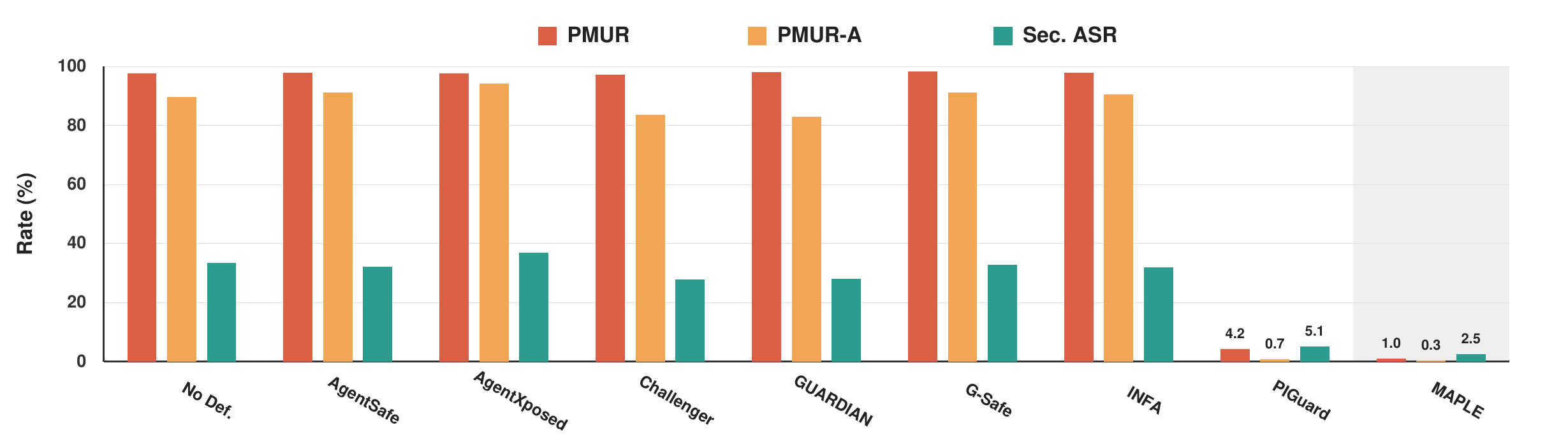}
\caption{Propagation metrics on AppWorld-AgentPoison. Bars report PMUR , PMUR-A , and Sec.\ ASR; definitions in Appendix~\ref{app:metrics}.}
\label{fig:propagation_mechanism}
\end{minipage}
\Description{Bar chart reporting poisoned-memory usage, usage by agents, and security-probe attack success for the evaluated guards on AppWorld-AgentPoison.}
\end{figure}

\paragraph{Obs3. One fixed guard configuration applies across the five evaluated attack settings.}
On Qwen3.5-122B-A10B, \methodname{} achieves the lowest mean RDA@3 in four settings, including 0.3\%, 0.9\%, and 0.2\% on MMLU, Long\-Mem\-Eval, and
  AppWorld, respectively. The signal functions, weights, and thresholds remain fixed without attack-specific parameter search. On AppWorld-AgentPoison, Figure~\ref{fig:propagation_mechanism} shows concurrent reductions in poisoned-memory use,
  agent-level exposure, and unsafe outputs. This association is consistent with blocked memory paths contributing to lower RDA, but does not rule out other differences between guard implementations. Section~\ref{sec:exp_utility} reports the corresponding attack-free utility results.

\begin{table}[!ht]
\centering
\footnotesize
\caption{PIGuard + full MAPLE on Gemma Base. Metrics (\%) are averaged equally over star, chain, and tree; bold marks the best mean. Appendix~\ref{app:ablation_protocol} details the protocol.}
\label{tab:piguard_composition_main}
\setlength{\tabcolsep}{2pt}
\begin{tabular*}{\columnwidth}{@{\extracolsep{\fill}}lrrrr@{}}
\toprule
& \multicolumn{2}{c}{\textbf{CSQA}} & \multicolumn{2}{c}{\textbf{LongMemEval}} \\
\cmidrule(lr){2-3}\cmidrule(lr){4-5}
Method & MDSR@3$\uparrow$ & RDA@3$\downarrow$ & MDSR@3$\uparrow$ & RDA@3$\downarrow$ \\
\midrule
PIGuard & 78.5 & 22.5 & 55.4 & 4.1 \\
MAPLE-Guard & 78.7 & 21.8 & 42.3 & \textbf{0.7} \\
PIGuard + MAPLE & \textbf{83.3} & \textbf{18.3} & \textbf{69.6} & 1.7 \\
\bottomrule
\end{tabular*}
\end{table}

Performance depends on benchmark and backbone (Table~\ref{tab:main_results}). MAPLE's Qwen AppWorld gains over PIGuard are small ($+0.4$ MDSR, $-0.5$ RDA points). PIGuard lowers both RDA and MDSR on Qwen CSQA and has higher MDSR on Gemma Long\-Mem\-Eval. On Gemma--InjectAgent, it leads both metrics (MDSR: 60.5\% versus 33.3\%; RDA: 32.3\% versus 56.1\%), exposing a limitation of MAPLE's current configuration. We additionally combine the same PIGuard retrieval filter with full MAPLE on Gemma (Table~\ref{tab:piguard_composition_main}; Appendix Table~\ref{tab:piguard_composition}). Relative to PIGuard, this raises MDSR and lowers RDA on CSQA ($+4.8/-4.2$ points) and Long\-Mem\-Eval ($+14.2/-2.4$ points). These means support additional benefits from the combined system in these two settings; the study jointly adds MAPLE's scoring, feedback, and gates. Paired uncertainty estimates remain necessary.
\subsection{Round-Level Propagation Dynamics}
\label{sec:exp_round}

\paragraph{Obs4. Safety degradation can accumulate across communication rounds.}
Figure~\ref{fig:roundwise_dynamics} traces round-level MDSR on MMLU, Long\-Mem\-Eval, and AppWorld under random communication topology; Appendix Tables~\ref{tab:mmlu_minja_round_appendix}--\ref{tab:round_injectagent} report the exact
  R1--R3 results for these settings and InjectAgent on both backbones; Appendix Table~\ref{tab:round_csqa_piguard} adds PIGuard on CSQA. From R1 to R3, MDSR under No Defense changes by \(-3.5\), \(-9.2\), and \(-19.5\) points on MMLU, Long\-Mem\-Eval, and AppWorld, respectively,
  compared with \(+4.5\), \(-4.4\), and \(0.0\) points under MAPLE-Guard. PIGuard changes by \(-2.7\), \(-1.1\), and \(0.0\) points, respectively, and matches MAPLE at 98.9\% on AppWorld. This pattern is consistent with repeated retrieval and cross-agent reuse amplifying stored influence and lifecycle filtering limiting
  exposure, but it does not establish causal mediation. These random-topology traces are stress-test results, not substitutes for the controlled-topology means in Table~\ref{tab:main_results}.

\begin{table}[!t]
\caption{Agent-scale retrieval-damage ASR across rounds.}
\label{tab:agent_scale_asr}
\centering
\footnotesize
\begin{tabular*}{\textwidth}{@{\extracolsep{\fill}}clccc@{}}
\toprule
\textbf{Agent Num.} & \textbf{Method} & \textbf{RDA@1}$\downarrow$ & \textbf{RDA@2}$\downarrow$ & \textbf{RDA@3}$\downarrow$ \\
\midrule
30 & No Defense & 21.7 & 24.5 & 26.5 \\
   & Ours       & 20.0 & 22.9 & 23.2 \\
\addlinespace
50 & No Defense & 20.5 & 22.1 & 24.3 \\
   & Ours       & 19.8 & 20.9 & 21.8 \\
\bottomrule
\end{tabular*}

\end{table}

\subsection{Topology and Scale Robustness}
\label{sec:exp_robust}

\paragraph{Obs5. \methodname{} maintains high mean MDSR and low mean RDA on AppWorld across controlled topologies.}
The star, chain, and tree settings vary the physical communication pattern while holding the task set fixed.
  Table~\ref{tab:main_results} summarizes the three topology scores using the reporting protocol in Section~5.1. On AppWorld-AgentPoison, MAPLE-Guard achieves 99.8\% MDSR@3 and 0.2\% RDA@3 with
  Qwen3.5-122B-A10B, and 95.3\% MDSR@3 and 0.0\% RDA@3 with gemma-4-31B. Appendix~\ref{app:pertopo} gives the topology breakdowns: Figure~\ref{fig:topology_breakdown_appworld} covers AppWorld, and Figures~\ref{fig:topology_breakdown_mmlu}--\ref{fig:topology_breakdown_injectagent} cover all benchmarks and backbones. Appendix~\ref{app:random} and Table~\ref{tab:random_topology_appendix} report the random-graph stress tests.

\paragraph{Obs6. The guard still reduces RDA in the larger-agent stress test.}
With fixed thresholds, we test 30 and 50 agents under random topologies (Table~\ref{tab:agent_scale_asr}). The backbone, attacker ratio, rounds, and graph policy are fixed; more agents also mean more attackers and interactions, so these are scale stress tests. At both scales, No Defense RDA rises from R1 to R3; MAPLE ends below it at 23.2\% and 21.8\% RDA@3, respectively. Its deterministic gates add no guard-LLM calls. Appendix~\ref{app:scale} details the controlled five-task token audit.

\begin{figure}[!t]
\centering
\begin{minipage}[t]{0.48\columnwidth}
\centering
\includegraphics[width=\linewidth]{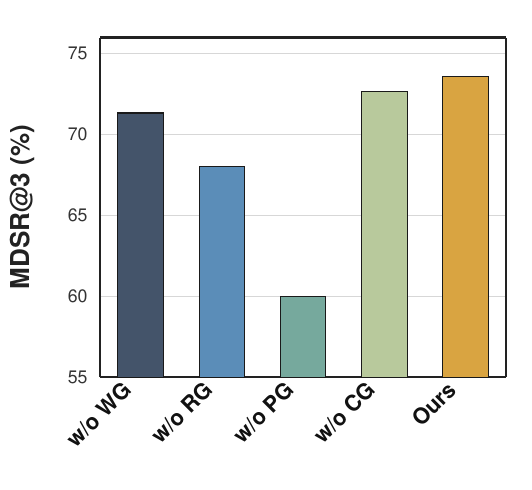}
\end{minipage}\hfill
\begin{minipage}[t]{0.48\columnwidth}
\centering
\includegraphics[width=\linewidth]{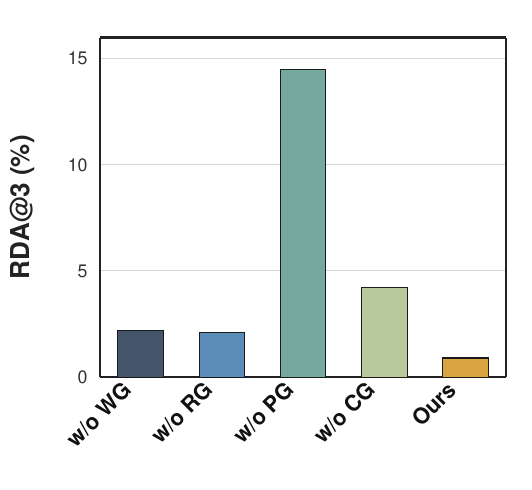}
\end{minipage}
\caption{Single-gate ablations on Long\-Mem\-Eval-MemoryGraft with full outcome feedback retained in every variant.}
\label{fig:ablation_components}
\Description{Two bar charts report MDSR and retrieval-damage ASR for full MAPLE-Guard and variants with one lifecycle gate removed on Long\-Mem\-Eval-MemoryGraft.}
\end{figure}

\subsection{Ablation Study}
\label{sec:exp_ablation}

We separate three questions: the contribution of individual gates, the effect of intervention across the memory lifecycle, and dependence on outcome feedback.

\begin{table}[t]
\centering
\small
\caption{Controlled ablations on the AppWorld action-selection proxy with Gemma Base. Metrics (\%) are equal-weight means over star, chain, tree, and random. Protocol and topology scores: Appendix~\ref{app:ablation_protocol}.}
\label{tab:controlled_ablations}
\setlength{\tabcolsep}{5pt}
\begin{tabular}{p{\dimexpr0.56\columnwidth-2\tabcolsep\relax}>{\raggedleft\arraybackslash}p{\dimexpr0.22\columnwidth-2\tabcolsep\relax}>{\raggedleft\arraybackslash}p{\dimexpr0.22\columnwidth-2\tabcolsep\relax}}
\toprule
\textbf{Variant} & \textbf{MDSR@3}$\uparrow$ & \textbf{RDA@3}$\downarrow$ \\
\midrule
\rowcolor{mapleguardgray}
\textbf{MAPLE-Guard (full)} & \textbf{88.4} & \textbf{0.1} \\
Retrieval-only & 65.0 & 21.6 \\
\texttt{no\_q}: w/o Q updates & 68.1 & 17.4 \\
\texttt{none}: w/o correctness feedback & 67.7 & 16.3 \\
\bottomrule
\end{tabular}
\end{table}

\paragraph{Obs7. Each lifecycle gate affects the measured defense.}
On Long\-Mem\-Eval-MemoryGraft, removing any single gate lowers MDSR@3 under the same outcome-conditioned update rule (Figure~\ref{fig:ablation_components}). The full guard achieves 73.6\% MDSR@3 and 0.9\% RDA@3. Removing the promotion gate causes the largest degradation, reducing MDSR@3 to 60.0\% and increasing RDA@3 to 14.5\%; removing the write, retrieval, or cross-agent gate produces smaller but consistent losses. This pattern is consistent with promotion being an important private-to-shared propagation point, as a contaminated item can become accessible to multiple agents after entering shared memory. These single-gate contrasts measure each component's contribution conditional on the remaining gates and full outcome feedback.

\paragraph{Matched lifecycle placement.}
A complementary controlled comparison uses the full guard and a retrieval-only variant with the same metadata access, risk scorer, and outcome-update rules. It tests the aggregate contribution of write, promotion, and cross-agent intervention beyond retrieval filtering. Each variant maintains its own evolving memory state under matched initial conditions and task order. Table~\ref{tab:controlled_ablations} reports 88.4\% MDSR@3 and 0.1\% RDA@3 for the full guard, compared with 65.0\% and 21.6\% for retrieval-only: a 23.4-point MDSR gain and a 21.6-point RDA reduction. Under this matched protocol, the results support intervention across the memory lifecycle beyond retrieval filtering.

\paragraph{Outcome-feedback dependence.}
The \texttt{no\_q} variant freezes outcome-driven Q updates while retaining the other feedback channels; \texttt{none} removes reference-correctness feedback from state updates, experience construction, and promotion eligibility. Both retain the lifecycle gates. Relative to the full guard, \texttt{no\_q} lowers MDSR@3 by 20.4 points and raises RDA@3 by 17.3 points; \texttt{none} lowers MDSR@3 by 20.8 points and raises RDA@3 by 16.2 points (Table~\ref{tab:controlled_ablations}). The full guard has the highest MDSR@3 and lowest RDA@3 in every topology (Appendix Table~\ref{tab:controlled_ablations_topology}). The two feedback variants have different metric rankings: \texttt{no\_q} has 0.4-point higher mean MDSR@3 but 1.1-point higher mean RDA@3 than \texttt{none}. Thus, task-level correctness and agent-output-level retrieval damage need not move together. Because \texttt{none} changes several feedback channels, its contrast measures their combined effect. These placement and feedback comparisons use the AppWorld action-selection proxy with Gemma Base across star, chain, tree, and random, with the full guard from the same protocol as the control. Appendix~\ref{app:ablation_protocol} details the interventions and reporting.

\subsection{Benign Utility and Token Usage}
\label{sec:exp_utility}

Table~\ref{tab:no_attack_qwen_random} brings the attack-free results into the main comparison. Under random topology, MAPLE changes MDSR by $-1.5$, $+7.0$, $+1.0$, $+1.4$, and $+0.9$ percentage points on the five Qwen settings, respectively. The corresponding gemma changes are $-1.5$, $-1.4$, $+5.3$, $+2.0$, and $+1.0$ points. Thus the guard is not uniformly cost-free: both backbones lose 1.5 points on MMLU, and gemma also loses 1.4 points on Long\-Mem\-Eval. These attack-free random-topology results characterize utility under their own protocol; they are not paired clean counterparts of the controlled-topology attack results in Table~\ref{tab:main_results}.

\begin{table}[t]
\caption{Attack-free MDSR@3 (\%) under random topology. Each cell shows No Defense $\rightarrow$ MAPLE. Qwen and gemma denote the same backbones as Table~\ref{tab:main_results}.}
\label{tab:no_attack_qwen_random}
\centering
\small
\begin{tabular*}{\columnwidth}{@{\extracolsep{\fill}}lcc@{}}
\toprule
Benchmark & Qwen & gemma \\
\midrule
MMLU & $88.5\rightarrow87.0$ & $56.5\rightarrow55.0$ \\
Long\-Mem\-Eval & $64.5\rightarrow71.5$ & $34.2\rightarrow32.8$ \\
AppWorld & $99.0\rightarrow100.0$ & $75.1\rightarrow80.4$ \\
CSQA & $80.1\rightarrow81.5$ & $81.5\rightarrow83.5$ \\
InjectAgent & $97.0\rightarrow97.9$ & $45.8\rightarrow46.8$ \\
\bottomrule
\end{tabular*}
\end{table}

The completed token audit uses five fixed tasks per benchmark, eight agents, three rounds, one fixed random adjacency and seed, and a generation limit of 128 tokens. MAPLE reduces measured token use relative to No Defense by 35.0\% on MMLU and 27.8\% on InjectAgent without additional guard-LLM calls (Appendix Figure~\ref{fig:token_overhead_random5}). This small audit measures token use in one configuration; it does not establish wall-clock speedup or replace the benign-utility comparison.

\FloatBarrier

\section{Conclusion}

  We formalize memory-link poisoning in LLM-based multi-agent systems and introduce MAPLE-Guard, a lifecycle-gated defense that treats memory as governed state at four checkpoints: write, retrieval, promotion, and
  cross-agent reuse. Across five benchmark--attack settings, multiple topologies, and two backbones, MAPLE-Guard reduces RDA and improves MDSR under attack. These results suggest that explicit memory-lifecycle
  controls can complement prompt- and topology-level safeguards.

\section*{Limitation}

  \methodname{} is designed to contain memory-link poisoning after it enters the system rather than prevent every initial compromise. It assumes reliable provenance metadata and retrieval traces; forged metadata or
  payloads designed to remain below per-event thresholds may evade the guard. The full-feedback setting uses benchmark-evaluator outcomes. Our controlled feedback variants distinguish outcome-driven Q updates from reference-correctness feedback, with their interpretation tied to the feedback source and benchmark protocol specified in Appendix~\ref{app:ablation_protocol}. The CSQA transfer runner also initializes part of its operational metadata from known attacker roles. Finally, because each benchmark is paired with a single attack, the evaluation supports within-benchmark comparisons among defenses but
  does not isolate the causal effect of the attack entry point.
\bibliography{upload/references}

@misc{altawaha2026remembering,
  title         = {Remembering More, Risking More: Longitudinal Safety Risks in Memory-Equipped LLM Agents},
  author        = {Al-Tawaha, Ahmad and Gu, Shangding and Niu, Peizhi and Jia, Ruoxi and Jin, Ming},
  year          = {2026},
  eprint        = {2605.17830},
  archivePrefix = {arXiv},
  primaryClass  = {cs.CR},
  url           = {https://arxiv.org/abs/2605.17830}
}

@misc{niu2026understanding,
  title         = {Understanding and Evaluating Claw-like Agent Security Through a Computer-Systems Lens},
  author        = {Niu, Peizhi and Qu, Wenjie and Gu, Shangding and Shi, Tianneng and Li, Yuankai and Tawaha, Ahmad and Alzahrani, Hend and Siu, Vincent and Li, Boyi and Wang, Chenguang and Zhang, Jiaheng and Alomair, Basel and Jin, Ming and Chen, Muhao and Wang, Chi and Spanos, Costas and Song, Dawn},
  year          = {2026},
  eprint        = {2606.30755},
  archivePrefix = {arXiv},
  primaryClass  = {cs.CR},
  url           = {https://arxiv.org/abs/2606.30755}
}

@article{zhong2024memorybank,
  title={MemoryBank: Enhancing Large Language Models with Long-Term Memory},
  author={Zhong, Wanjun and Guo, Lianghong and Gao, Qiqi and Ye, He and Wang, Yanlin},
  journal={Proceedings of the AAAI Conference on Artificial Intelligence},
  year={2024}
}

@inproceedings{greshake2023indirect,
  title={Not What You've Signed Up For: Compromising Real-World LLM-Integrated Applications with Indirect Prompt Injection},
  author={Greshake, Kai and Abdelnabi, Sahar and Mishra, Shailesh and Endres, Christoph and Holz, Thorsten and Fritz, Mario},
  booktitle={Proceedings of the 16th ACM Workshop on Artificial Intelligence and Security},
  year={2023},
  doi={10.1145/3605764.3623985}
}

@inproceedings{zhong2023poisoning,
  title={Poisoning Retrieval Corpora by Injecting Adversarial Passages},
  author={Zhong, Zexuan and Huang, Ziqing and Wettig, Alexander and Chen, Danqi},
  booktitle={Proceedings of the 2023 Conference on Empirical Methods in Natural Language Processing},
  pages={13764--13775},
  year={2023},
  doi={10.18653/v1/2023.emnlp-main.849}
}

@inproceedings{zou2025poisonedrag,
  title={PoisonedRAG: Knowledge Corruption Attacks to Retrieval-Augmented Generation of Large Language Models},
  author={Zou, Wei and Geng, Runpeng and Wang, Binghui and Jia, Jinyuan},
  booktitle={34th USENIX Security Symposium},
  year={2025}
}

@article{packer2023memgpt,
  title={MemGPT: Towards LLMs as Operating Systems},
  author={Packer, Charles and Fang, Vivian and Patil, Shishir G. and Lin, Kevin and Wooders, Sarah and Gonzalez, Joseph E.},
  journal={arXiv preprint arXiv:2310.08560},
  year={2023}
}

@article{chhikara2025mem0,
  title={Mem0: Building Production-Ready AI Agents with Scalable Long-Term Memory},
  author={Chhikara, Prateek and Khant, Dev and Aryan, Saket and Singh, Taranjeet and Yadav, Deshraj},
  journal={arXiv preprint arXiv:2504.19413},
  year={2025}
}

@inproceedings{park2023generative,
  title={Generative Agents: Interactive Simulacra of Human Behavior},
  author={Park, Joon Sung and O'Brien, Joseph C. and Cai, Carrie J. and Morris, Meredith Ringel and Liang, Percy and Bernstein, Michael S.},
  booktitle={Proceedings of the 36th Annual ACM Symposium on User Interface Software and Technology},
  year={2023}
}

@inproceedings{shinn2023reflexion,
  title     = {Reflexion: Language Agents with Verbal Reinforcement Learning},
  author    = {Shinn, Noah and Cassano, Federico and Gopinath, Ashwin
               and Narasimhan, Karthik and Yao, Shunyu},
  booktitle = {Advances in Neural Information Processing Systems},
  editor    = {Oh, Alice and Naumann, Tristan and Globerson, Amir
               and Saenko, Kate and Hardt, Moritz and Levine, Sergey},
  volume    = {36},
  pages     = {8634--8652},
  publisher = {Curran Associates, Inc.},
  year      = {2023},
  url       = {https://proceedings.neurips.cc/paper_files/paper/2023/file/1b44b878bb782e6954cd888628510e90-Paper-Conference.pdf}
}

@article{wang2024voyager,
  title   = {Voyager: An Open-Ended Embodied Agent with Large Language Models},
  author  = {Wang, Guanzhi and Xie, Yuqi and Jiang, Yunfan
             and Mandlekar, Ajay and Xiao, Chaowei and Zhu, Yuke
             and Fan, Linxi and Anandkumar, Anima},
  journal = {Transactions on Machine Learning Research},
  year    = {2024},
  url     = {https://openreview.net/forum?id=ehfRiF0R3a}
}

@inproceedings{maharana2024locomo,
  title={Evaluating Very Long-Term Conversational Memory of LLM Agents},
  author={Maharana, Adyasha and Lee, Dong-Ho and Tulyakov, Sergey and Bansal, Mohit and Barbieri, Francesco and Fang, Yuwei},
  booktitle={Proceedings of the 62nd Annual Meeting of the Association for Computational Linguistics},
  year={2024}
}

@inproceedings{wu2025longmemeval,
  title     = {{LongMemEval}: Benchmarking Chat Assistants on Long-Term Interactive Memory},
  author    = {Wu, Di and Wang, Hongwei and Yu, Wenhao
               and Zhang, Yuwei and Chang, Kai-Wei and Yu, Dong},
  booktitle = {International Conference on Learning Representations},
  editor    = {Yue, Yisong and Garg, Animesh and Peng, Nanyun
               and Sha, Fei and Yu, Rose},
  volume    = {2025},
  pages     = {86809--86836},
  year      = {2025},
  url       = {https://proceedings.iclr.cc/paper_files/paper/2025/file/d813d324dbf0598bbdc9c8e79740ed01-Paper-Conference.pdf}
}

@inproceedings{xu2025amem,
  title     = {{A-MEM}: Agentic Memory for {LLM} Agents},
  author    = {Xu, Wujiang and Liang, Zujie and Mei, Kai
               and Gao, Hang and Tan, Juntao and Zhang, Yongfeng},
  booktitle = {Advances in Neural Information Processing Systems},
  editor    = {Belgrave, Danielle and Zhang, Cheng and Lin, Hsuan-Tien
               and Pascanu, Razvan and Koniusz, Piotr
               and Ghassemi, Marzyeh and Chen, Nan},
  volume    = {38},
  pages     = {17577--17604},
  publisher = {Curran Associates, Inc.},
  year      = {2025},
  url       = {https://proceedings.neurips.cc/paper_files/paper/2025/file/19909c36f51abc4856b4560aff3d36d6-Paper-Conference.pdf}
}

@article{zhou2025memento,
  title={Memento: Fine-tuning LLM Agents without Fine-tuning LLMs},
  author={Zhou, Huichi and Chen, Yihang and Guo, Siyuan and Yan, Xue and Lee, Kin Hei and Wang, Zihan and Lee, Ka Yiu and Zhang, Guchun and Shao, Kun and Yang, Linyi and Wang, Jun},
  journal={arXiv preprint arXiv:2508.16153},
  year={2025}
}

@article{memrl2026,
  title={MemRL: Self-Evolving Agents via Runtime Reinforcement Learning on Episodic Memory},
  author={Zhang, Shengtao and Wang, Jiaqian and Zhou, Ruiwen and Liao, Junwei and Feng, Yuchen and Li, Zhuo and Zheng, Yujie and Zhang, Weinan and Wen, Ying and Li, Zhiyu and Xiong, Feiyu and Qi, Yutao and Tang, Bo and Wen, Muning},
  journal={arXiv preprint arXiv:2601.03192},
  year={2026}
}

@inproceedings{dong2025memory,
  title     = {Memory Injection Attacks on {LLM} Agents via Query-Only Interaction},
  author    = {Dong, Shen and Xu, Shaochen and He, Pengfei and Li, Yige
               and Tang, Jiliang and Liu, Tianming and Liu, Hui and Xiang, Zhen},
  booktitle = {Advances in Neural Information Processing Systems},
  editor    = {Belgrave, Danielle and Zhang, Cheng and Lin, Hsuan-Tien
               and Pascanu, Razvan and Koniusz, Piotr and Ghassemi, Marzyeh
               and Chen, Nan},
  volume    = {38},
  pages     = {46697--46731},
  publisher = {Curran Associates, Inc.},
  year      = {2025},
  url       = {https://proceedings.neurips.cc/paper_files/paper/2025/file/42a97bbd9844d2bf68596730af80bcdf-Paper-Conference.pdf}
}

@inproceedings{chen2024agentpoison,
  title={AgentPoison: Red-teaming LLM Agents via Poisoning Memory or Knowledge Bases},
  author={Chen, Zhaorun and Xiang, Zhen and Xiao, Chaowei and Song, Dawn and Li, Bo},
  booktitle={Advances in Neural Information Processing Systems},
  year={2024}
}

@article{srivastava2025memorygraft,
  title={MemoryGraft: Persistent Compromise of LLM Agents via Poisoned Experience Retrieval},
  author={Srivastava, Saksham Sahai and He, Haoyu},
  journal={arXiv preprint arXiv:2512.16962},
  year={2025}
}

@inproceedings{li2023camel,
  title={{CAMEL}: Communicative Agents for ``Mind'' Exploration of Large Language Model Society},
  author={Li, Guohao and Hammoud, Hasan Abed Al Kader and Itani, Hani and Khizbullin, Dmitrii and Ghanem, Bernard},
  booktitle={Advances in Neural Information Processing Systems},
  year={2023}
}

@article{wu2023autogen,
  title={AutoGen: Enabling Next-Gen LLM Applications via Multi-Agent Conversation},
  author={Wu, Qingyun and Bansal, Gagan and Zhang, Jieyu and Wu, Yiran and Li, Beibin and Zhu, Erkang and Jiang, Li and Zhang, Xiaoyun and Zhang, Shaokun and Liu, Jiale and Awadallah, Ahmed Hassan and White, Ryen W. and Burger, Doug and Wang, Chi},
  journal={arXiv preprint arXiv:2308.08155},
  year={2023}
}

@inproceedings{hong2024metagpt,
  title={Meta{GPT}: Meta Programming for A Multi-Agent Collaborative Framework},
  author={Hong, Sirui and Zhuge, Mingchen and Chen, Jonathan and Zheng, Xiawu and Cheng, Yuheng and Wang, Jinlin and Zhang, Ceyao and Wang, Zili and Yau, Steven Ka Shing and Lin, Zijuan and Zhou, Liyang and Ran, Chenyu and Xiao, Lingfeng and Wu, Chenglin and Schmidhuber, J{\"u}rgen},
  booktitle={International Conference on Learning Representations},
  year={2024}
}

@inproceedings{qian2024chatdev,
  title={ChatDev: Communicative Agents for Software Development},
  author={Qian, Chen and Liu, Wei and Liu, Hongzhang and Chen, Nuo and Dang, Yufan and Li, Jiahao and Yang, Cheng and Chen, Weize and Su, Yusheng and Cong, Xin and Xu, Juyuan and Li, Dahai and Liu, Zhiyuan and Sun, Maosong},
  booktitle={Proceedings of the 62nd Annual Meeting of the Association for Computational Linguistics},
  year={2024}
}

@article{torra2026memorypoisoningmas,
  title={Memory Poisoning and Secure Multi-Agent Systems},
  author={Torra, Vicen{\c{c}} and Bras-Amor{\'o}s, Maria},
  journal={arXiv preprint arXiv:2603.20357},
  year={2026}
}

@inproceedings{wang2025gsafeguard,
  title={{G-Safeguard}: A Topology-Guided Security Lens and Treatment on LLM-based Multi-agent Systems},
  author={Wang, Shilong and Zhang, Guibin and Yu, Miao and Wan, Guancheng and Meng, Fanci and Guo, Chongye and Wang, Kun and Wang, Yang},
  booktitle={Proceedings of the 63rd Annual Meeting of the Association for Computational Linguistics},
  year={2025}
}

@article{zhou2026infaguard,
  title={{INFA-Guard}: Mitigating Malicious Propagation via Infection-Aware Safeguarding in LLM-Based Multi-Agent Systems},
  author={Zhou, Yijin and Lu, Xiaoya and Liu, Dongrui and Yan, Junchi and Shao, Jing},
  journal={arXiv preprint arXiv:2601.14667},
  year={2026}
}

@inproceedings{xiang2025guardagent,
  title     = {{GuardAgent}: Safeguard {LLM} Agents via Knowledge-Enabled Reasoning},
  author    = {Xiang, Zhen and Zheng, Linzhi and Li, Yanjie
               and Hong, Junyuan and Li, Qinbin and Xie, Han
               and Zhang, Jiawei and Xiong, Zidi and Xie, Chulin
               and Yang, Carl and Song, Dawn and Li, Bo},
  booktitle = {Proceedings of the 42nd International Conference on Machine Learning},
  editor    = {Singh, Aarti and Fazel, Maryam and Hsu, Daniel
               and Lacoste-Julien, Simon and Berkenkamp, Felix
               and Maharaj, Tegan and Wagstaff, Kiri and Zhu, Jerry},
  series    = {Proceedings of Machine Learning Research},
  volume    = {267},
  pages     = {68316--68342},
  publisher = {PMLR},
  year      = {2025},
  url       = {https://proceedings.mlr.press/v267/xiang25a.html}
}

@article{mao2025agentsafe,
  title={AgentSafe: Safeguarding Large Language Model-based Multi-agent Systems via Hierarchical Data Management},
  author={Mao, Junyuan and Meng, Fanci and Duan, Yifan and Yu, Miao and Jia, Xiaojun and Fang, Junfeng and Liang, Yuxuan and Wang, Kun and Wen, Qingsong},
  journal={arXiv preprint arXiv:2503.04392},
  year={2025}
}

@article{xie2025agentxposed,
  title={Who's the Mole? Modeling and Detecting Intention-Hiding Malicious Agents in LLM-Based Multi-Agent Systems},
  author={Xie, Yizhe and Zhu, Congcong and Zhang, Xinyue and Zhu, Tianqing and Ye, Dayong and Wang, Minghao and Liu, Chi},
  journal={arXiv preprint arXiv:2507.04724},
  year={2025}
}

@inproceedings{huang2025resilience,
  title={On the Resilience of LLM-Based Multi-Agent Collaboration with Faulty Agents},
  author={Huang, Jen-Tse and Zhou, Jiaxu and Jin, Tailin and Zhou, Xuhui and Chen, Zixi and Wang, Wenxuan and Yuan, Youliang and Lyu, Michael and Sap, Maarten},
  booktitle={Proceedings of the 42nd International Conference on Machine Learning},
  year={2025}
}

@inproceedings{hendrycks2021mmlu,
  title={Measuring Massive Multitask Language Understanding},
  author={Hendrycks, Dan and Burns, Collin and Basart, Steven and Zou, Andy and Mazeika, Mantas and Song, Dawn and Steinhardt, Jacob},
  booktitle={International Conference on Learning Representations},
  year={2021}
}

@inproceedings{talmor2019commonsenseqa,
  title={CommonsenseQA: A Question Answering Challenge Targeting Commonsense Knowledge},
  author={Talmor, Alon and Herzig, Jonathan and Lourie, Nicholas and Berant, Jonathan},
  booktitle={Proceedings of the 2019 Conference of the North American Chapter of the Association for Computational Linguistics},
  pages={4149--4158},
  year={2019}
}

@inproceedings{trivedi2024appworld,
  title={AppWorld: A Controllable World of Apps and People for Benchmarking Interactive Coding Agents},
  author={Trivedi, Harsh and Khot, Tushar and Hartmann, Mareike and Manku, Ruskin and Dong, Vinty and Li, Edward and Gupta, Shashank and Sabharwal, Ashish and Balasubramanian, Niranjan},
  booktitle={Proceedings of the 62nd Annual Meeting of the Association for Computational Linguistics},
  pages={16022--16076},
  year={2024},
  doi={10.18653/v1/2024.acl-long.850}
}

@inproceedings{zhan2024injecagent,
  title={InjecAgent: Benchmarking Indirect Prompt Injections in Tool-Integrated Large Language Model Agents},
  author={Zhan, Qiusi and Liang, Zhixiang and Ying, Zifan and Kang, Daniel},
  booktitle={Findings of the Association for Computational Linguistics: ACL 2024},
  pages={10471--10506},
  year={2024},
  doi={10.18653/v1/2024.findings-acl.624}
}

@article{watkins1992qlearning,
  title={Q-learning},
  author={Watkins, Christopher J. C. H. and Dayan, Peter},
  journal={Machine Learning},
  volume={8},
  number={3--4},
  pages={279--292},
  year={1992}
}

@book{sutton2018reinforcement,
  title={Reinforcement Learning: An Introduction},
  author={Sutton, Richard S. and Barto, Andrew G.},
  edition={2},
  publisher={MIT Press},
  year={2018}
}

@inproceedings{yao2023react,
  title={{ReAct}: Synergizing Reasoning and Acting in Language Models},
  author={Yao, Shunyu and Zhao, Jeffrey and Yu, Dian and Du, Nan and Shafran, Izhak and Narasimhan, Karthik and Cao, Yuan},
  booktitle={International Conference on Learning Representations},
  year={2023}
}

@inproceedings{schick2023toolformer,
  title={Toolformer: Language Models Can Teach Themselves to Use Tools},
  author={Schick, Timo and Dwivedi-Yu, Jane and Dess{\`\i}, Roberto and Raileanu, Roberta and Lomeli, Maria and Zettlemoyer, Luke and Cancedda, Nicola and Scialom, Thomas},
  booktitle={Advances in Neural Information Processing Systems},
  year={2023}
}

@article{inan2023llamaguard,
  title={Llama Guard: {LLM}-based Input-Output Safeguard for Human-{AI} Conversations},
  author={Inan, Hakan and Upasani, Kartikeya and Chi, Jianfeng and Rungta, Rashi and Iyer, Krithika and Mao, Yuning and Tontchev, Michael and Hu, Qing and Fuller, Brian and Testuggine, Davide and Khabsa, Madian},
  journal={arXiv preprint arXiv:2312.06674},
  year={2023}
}

@inproceedings{lewis2020rag,
  title={Retrieval-Augmented Generation for Knowledge-Intensive {NLP} Tasks},
  author={Lewis, Patrick and Perez, Ethan and Piktus, Aleksandra and Petroni, Fabio and Karpukhin, Vladimir and Goyal, Naman and K{\"u}ttler, Heinrich and Lewis, Mike and Yih, Wen-tau and Rockt{\"a}schel, Tim and Riedel, Sebastian and Kiela, Douwe},
  booktitle={Advances in Neural Information Processing Systems},
  year={2020}
}

@article{denning1976lattice,
  title={A Lattice Model of Secure Information Flow},
  author={Denning, Dorothy E.},
  journal={Communications of the ACM},
  volume={19},
  number={5},
  pages={236--243},
  year={1976},
  publisher={ACM}
}

@inproceedings{newsome2005dynamic,
  title={Dynamic Taint Analysis for Automatic Detection, Analysis, and Signature Generation of Exploits on Commodity Software},
  author={Newsome, James and Song, Dawn},
  booktitle={Proceedings of the Network and Distributed System Security Symposium (NDSS)},
  year={2005}
}

@misc{lin2026memorylifecycle,
  title={A Survey on Long-Term Memory Security in {LLM} Agents: Attacks, Defenses, and Governance Across the Memory Lifecycle},
  author={Lin, Zehao and Hao, Xixuan and Fu, Renyu and Cui, Shaobo and Chen, Kai and Li, Chunyu and Li, Zhiyu and Xiong, Feiyu},
  year={2026},
  eprint={2604.16548},
  archivePrefix={arXiv},
  primaryClass={cs.CR}
}

@misc{sharma2026smsr,
  title={{SMSR}: Certified Defence Against Runtime Memory Poisoning in Persistent {LLM} Agent Systems},
  author={Sharma, Tarun},
  year={2026},
  eprint={2606.12703},
  archivePrefix={arXiv},
  primaryClass={cs.CR}
}

@misc{dash2026untrusted,
  title={From Untrusted Input to Trusted Memory: A Systematic Study of Memory Poisoning Attacks in {LLM} Agents},
  author={Dash, Pritam and Ge, Tongyu and Jain, Aditi and Shah, Tanmay and Shang, Zhiwei},
  year={2026},
  eprint={2606.04329},
  archivePrefix={arXiv},
  primaryClass={cs.CR}
}

@inproceedings{pan2026xgguard,
    title = "Explainable and Fine-Grained Safeguarding of {LLM} Multi-Agent Systems via Bi-Level Graph Anomaly Detection",
    author = "Pan, Junjun  and
      Liu, Yixin  and
      Miao, Rui  and
      Ding, Kaize  and
      Zheng, Yu  and
      Nguyen, Quoc Viet Hung  and
      Liew, Alan Wee-Chung  and
      Pan, Shirui",
    editor = "Liakata, Maria  and
      Moreira, Viviane P.  and
      Zhang, Jiajun  and
      Jurgens, David",
    booktitle = "Proceedings of the 64th Annual Meeting of the {A}ssociation for {C}omputational {L}inguistics (Volume 1: Long Papers)",
    month = jul,
    year = "2026",
    address = "San Diego, California, United States",
    publisher = "Association for Computational Linguistics",
    url = "https://aclanthology.org/2026.acl-long.1407/",
    doi = "10.18653/v1/2026.acl-long.1407",
    pages = "30491--30506",
    ISBN = "979-8-89176-390-6"
}

@inproceedings{li2025piguard,
    title = "{PIG}uard: Prompt Injection Guardrail via Mitigating Overdefense for Free",
    author = "Li, Hao  and
      Liu, Xiaogeng  and
      Zhang, Ning  and
      Xiao, Chaowei",
    editor = "Che, Wanxiang  and
      Nabende, Joyce  and
      Shutova, Ekaterina  and
      Pilehvar, Mohammad Taher",
    booktitle = "Proceedings of the 63rd Annual Meeting of the Association for Computational Linguistics (Volume 1: Long Papers)",
    month = jul,
    year = "2025",
    address = "Vienna, Austria",
    publisher = "Association for Computational Linguistics",
    url = "https://aclanthology.org/2025.acl-long.1468/",
    doi = "10.18653/v1/2025.acl-long.1468",
    pages = "30420--30437",
    ISBN = "979-8-89176-251-0"
}

@inproceedings{li2025drift,
 author = {Li, Hao and Liu, Xiaogeng and Chiu, Hung-Chun and Li, Dianqi and Zhang, Ning and Xiao, Chaowei},
 booktitle = {Advances in Neural Information Processing Systems},
 doi = {10.52202/085713-2791},
 editor = {D. Belgrave and C. Zhang and H. Lin and R. Pascanu and P. Koniusz and M. Ghassemi and N. Chen},
 pages = {83262--83290},
 publisher = {Curran Associates, Inc.},
 title = {DRIFT: Dynamic Rule-Based Defense with Injection Isolation for Securing LLM Agents},
 url = {https://proceedings.neurips.cc/paper_files/paper/2025/file/77f3b26c7907aa27b207df9b9d43f29a-Paper-Conference.pdf},
 volume = {38},
 year = {2025}
}

@inproceedings{debenedetti2024agentdojo,
 author = {Debenedetti, Edoardo and Zhang, Jie and Balunovic, Mislav and Beurer-Kellner, Luca and Fischer, Marc and Tram\`{e}r, Florian},
 booktitle = {Advances in Neural Information Processing Systems},
 doi = {10.52202/079017-2636},
 editor = {A. Globerson and L. Mackey and D. Belgrave and A. Fan and U. Paquet and J. Tomczak and C. Zhang},
 pages = {82895--82920},
 publisher = {Curran Associates, Inc.},
 title = {AgentDojo: A Dynamic Environment to Evaluate Prompt Injection Attacks and Defenses for LLM Agents},
 url = {https://proceedings.neurips.cc/paper_files/paper/2024/file/97091a5177d8dc64b1da8bf3e1f6fb54-Paper-Datasets_and_Benchmarks_Track.pdf},
 volume = {37},
 year = {2024}
}

@misc{debenedetti2025camel,
  title={Defeating Prompt Injections by Design},
  author={Debenedetti, Edoardo and Shumailov, Ilia and Fan, Tianqi and Hayes, Jamie and Carlini, Nicholas and Fabian, Daniel and Kern, Christoph and Shi, Chongyang and Terzis, Andreas and Tram{\`e}r, Florian},
  year={2025},
  eprint={2503.18813},
  archivePrefix={arXiv},
  primaryClass={cs.CR},
  url={https://arxiv.org/abs/2503.18813}
}

@inproceedings{zhang2025asb,
  title={Agent Security Bench ({ASB}): Formalizing and Benchmarking Attacks and Defenses in {LLM}-based Agents},
  author={Zhang, Hanrong and Huang, Jingyuan and Mei, Kai and Yao, Yifei and Wang, Zhenting and Zhan, Chenlu and Wang, Hongwei and Zhang, Yongfeng},
  booktitle={International Conference on Learning Representations},
  year={2025},
  url={https://openreview.net/forum?id=V4y0CpX4hK}
}

@inproceedings{chen2025struq,
  title={{StruQ}: Defending Against Prompt Injection with Structured Queries},
  author={Chen, Sizhe and Piet, Julien and Sitawarin, Chawin and Wagner, David},
  booktitle={34th USENIX Security Symposium (USENIX Security 25)},
  publisher={USENIX Association},
  year={2025},
  pages={2383--2400},
  url={https://www.usenix.org/conference/usenixsecurity25/presentation/chen-sizhe}
}

@inproceedings{liu2025datasentinel,
  title={{DataSentinel}: A Game-Theoretic Detection of Prompt Injection Attacks},
  author={Liu, Yupei and Jia, Yuqi and Jia, Jinyuan and Song, Dawn and Gong, Neil Zhenqiang},
  booktitle={2025 IEEE Symposium on Security and Privacy (SP)},
  year={2025},
  url={https://arxiv.org/abs/2504.11358}
}

@inproceedings{wei2026amemguard,
  title = {{A-MemGuard}: A Proactive Defense Framework For {LLM}-Based Agent Memory},
  author = {Wei, Qianshan and Yang, Tengchao and Wang, Yaochen and Li, Xinfeng and Li, Lijun and Yin, Zhenfei and Zhan, Yi and Holz, Thorsten and Lin, Zhiqiang and Wang, Xiaofeng},
  booktitle = {Proceedings of the 43rd International Conference on Machine Learning},
  series = {Proceedings of Machine Learning Research},
  volume = {306},
  pages = {134176--134202},
  publisher = {PMLR},
  year = {2026},
  url = {https://proceedings.mlr.press/v306/wei26ac.html}
}

@inproceedings{kang2025memoryos,
  title = {Memory {OS} of {AI} Agent},
  author = {Kang, Jiazheng and Ji, Mingming and Zhao, Zhe and Bai, Ting},
  booktitle = {Proceedings of the 2025 Conference on Empirical Methods in Natural Language Processing},
  pages = {25961--25970},
  address = {Suzhou, China},
  publisher = {Association for Computational Linguistics},
  year = {2025},
  doi = {10.18653/v1/2025.emnlp-main.1318},
  url = {https://aclanthology.org/2025.emnlp-main.1318/}
}

@inproceedings{hu2026memoryagentbench,
  title = {Evaluating Memory in {LLM} Agents via Incremental Multi-Turn Interactions},
  author = {Hu, Yuanzhe and Wang, Yu and McAuley, Julian},
  booktitle = {The Fourteenth International Conference on Learning Representations},
  year = {2026},
  url = {https://openreview.net/forum?id=DT7JyQC3MR}
}

@inproceedings{gu2024agentsmith,
  title = {Agent Smith: A Single Image Can Jailbreak One Million Multimodal {LLM} Agents Exponentially Fast},
  author = {Gu, Xiangming and Zheng, Xiaosen and Pang, Tianyu and Du, Chao and Liu, Qian and Wang, Ye and Jiang, Jing and Lin, Min},
  booktitle = {Proceedings of the 41st International Conference on Machine Learning},
  series = {Proceedings of Machine Learning Research},
  volume = {235},
  pages = {16647--16672},
  publisher = {PMLR},
  year = {2024},
  url = {https://proceedings.mlr.press/v235/gu24e.html}
}

@misc{lee2024promptinfection,
  title = {Prompt Infection: {LLM}-to-{LLM} Prompt Injection within Multi-Agent Systems},
  author = {Lee, Donghyun and Tiwari, Mo},
  year = {2024},
  eprint = {2410.07283},
  archivePrefix = {arXiv},
  url = {https://arxiv.org/abs/2410.07283}
}

@misc{wallace2024instructionhierarchy,
  title = {The Instruction Hierarchy: Training {LLMs} to Prioritize Privileged Instructions},
  author = {Wallace, Eric and Xiao, Kai and Leike, Reimar and Weng, Lilian and Heidecke, Johannes and Beutel, Alex},
  year = {2024},
  eprint = {2404.13208},
  archivePrefix = {arXiv},
  url = {https://arxiv.org/abs/2404.13208}
}

@misc{hines2024spotlighting,
  title = {Defending Against Indirect Prompt Injection Attacks With Spotlighting},
  author = {Hines, Keegan and Lopez, Gary and Hall, Matthew and Zarfati, Federico and Zunger, Yonatan and Kiciman, Emre},
  year = {2024},
  eprint = {2403.14720},
  archivePrefix = {arXiv},
  url = {https://arxiv.org/abs/2403.14720}
}

@inproceedings{chen2025secalign,
  title = {{SecAlign}: Defending Against Prompt Injection with Preference Optimization},
  author = {Chen, Sizhe and Zharmagambetov, Arman and Mahloujifar, Saeed and Chaudhuri, Kamalika and Wagner, David and Guo, Chuan},
  booktitle = {Proceedings of the 2025 ACM SIGSAC Conference on Computer and Communications Security},
  publisher = {Association for Computing Machinery},
  year = {2025},
  url = {https://sizhe-chen.github.io/SecAlign-Website/}
}

@inproceedings{yi2025bipia,
  title = {Benchmarking and Defending Against Indirect Prompt Injection Attacks on Large Language Models},
  author = {Yi, Jingwei and Xie, Yueqi and Zhu, Bin and Kiciman, Emre and Sun, Guangzhong and Xie, Xing and Wu, Fangzhao},
  booktitle = {Proceedings of the 31st ACM SIGKDD Conference on Knowledge Discovery and Data Mining},
  publisher = {Association for Computing Machinery},
  year = {2025},
  doi = {10.1145/3690624.3709179},
  url = {https://arxiv.org/abs/2312.14197}
}

@inproceedings{zhan2025adaptiveattacks,
  title = {Adaptive Attacks Break Defenses Against Indirect Prompt Injection Attacks on {LLM} Agents},
  author = {Zhan, Qiusi and Fang, Richard and Panchal, Henil Shalin and Kang, Daniel},
  booktitle = {Findings of the Association for Computational Linguistics: NAACL 2025},
  publisher = {Association for Computational Linguistics},
  pages = {7116--7132},
  year = {2025},
  url = {https://aclanthology.org/2025.findings-naacl.395/}
}
\clearpage
\appendix
% Allow modest extra interword stretch only within the appendix when a line is otherwise too long.
\begingroup
\setlength{\emergencystretch}{2em}
\section{Additional Method Details}
\label{app:method}

This section gives the implementation-level details that are abbreviated in the main text (Sections~\ref{sec:threat} and~\ref{sec:mapleguard}). The central object is not a message edge in the communication graph, but a memory-link event: an operation that writes, retrieves, promotes, reuses, or updates a memory item. A memory-link path is the ordered sequence of such events for the same item. The path may cross agents through shared memory even when no direct message is sent between the agents.

\subsection{Memory State and Hidden Signals}

Each memory item contains text and metadata. The backbone receives the memory text and, in the current MAPLE serialization, may also receive a limited rendering of operational origin fields; it does not receive the full guard state, provenance chain, or hidden risk labels. The metadata records scope, provenance, lifecycle state, and guard-side signals. Scope determines whether the item is private, shared, blocked, or quarantined. Provenance records where the item came from. The lifecycle state records whether the item can be retrieved or promoted. The hidden guard-side signals include risk, source trust, retrieval utility, and outcome history. These signals are used by the guard but are not exposed as instructions to the backbone LLM.

\subsection{Runtime Metadata and Evaluator Separation}

The memory manager records defense-visible metadata when an operation occurs: writer/origin ID, source route (e.g., agent output, external input, tool observation, or imported memory), requested scope, access policy, verification state, parent memory IDs, and subsequent success/failure counts. Writer ID is an operational identifier, not an attacker flag. Scope and lineage follow from write, promotion, and cross-agent events; content hazard and query-conditioned signals are deterministic functions of this metadata, the memory text, and the current query. For the primary memory-native paths, the field called \emph{taint} is an operational state such as clean, unverified, or external rather than the evaluator's ``poisoned'' ground-truth label; runner-specific role-dependent initialization is described below.

For the primary memory-native paths, attack names are removed from defense-visible source strings and parent markers, and poison IDs, the target $y^\star$, and attack-success annotations remain in runner-side evaluator maps used to compute RDA, MDSR, PMUR, and related metrics. They are not separately supplied to the backbone as evaluator annotations. In the full-feedback configuration, the benchmark evaluator uses reference answers after each task to derive the success/failure signal for memory utility and provenance-trust updates. The controlled feedback study distinguishes this configuration from frozen outcome-driven Q updates (\texttt{no\_\allowbreak{}q}) and removal of reference-correctness feedback (\texttt{none}); Section~\ref{app:ablation_protocol} specifies which channels remain active in each variant. Reference answers remain available to the evaluator for metric computation in all variants. Deployment interpretation follows the feedback source used by each configuration. In the CSQA transfer runner, preassigned attacker roles also initialize part of the operational metadata; we therefore treat CSQA as a transfer probe rather than evidence that attacker identity is available in deployment.

The evaluator separation keeps the evaluator's attack-target annotation from being separately supplied to the backbone, but it does not eliminate these evaluator-assisted state updates; the attack payload itself can contain a target answer or action. The guard also assumes that runtime metadata itself is trustworthy; the current evaluation does not cover an attacker who forges source-route, writer, scope, or lineage records.

\subsection{Lifecycle Gates}

\methodname{} applies four gates to the memory lifecycle. The write gate rejects or quarantines unsafe memory candidates before they become persistent state. The retrieval gate filters memories before they enter the reasoning context. The promotion gate prevents contaminated private memories from entering shared memory. The cross-agent gate blocks unsafe reuse when a memory item is about to influence another agent. The gates share the same metadata, so a memory item that is suspicious at write time can remain restricted during retrieval and promotion.

\subsection{Implementation and Hyperparameters}

\methodname{} uses a single fixed configuration across all five evaluated benchmarks and attacks. The scorer makes no direct call to an LLM or trained classifier and receives no task answer or target label as a scoring argument; its state can nevertheless reflect evaluator-derived outcome updates and runner-specific role metadata, as described above. Its direct inputs are deterministic provenance, metadata, token/regex, and store-statistics functions, including fixed steering and trigger phrase lists, and fixed weights and gate cutoffs. No attack-name selector is passed to the scorer. Table~\ref{tab:hyperparams} lists the values. The implementation specifies these parameters as rule constants rather than learning them in the scorer. The configuration gives provenance and lineage the largest combined weight and applies a lower total-risk cutoff to team promotion than to retrieval. These design choices do not establish numerical optimality or robustness to parameter changes. The unified risk of Eq.~\eqref{eq:risk} places the largest weight on provenance and lineage; the retrieval score of Eq.~\eqref{eq:retrieval} rewards utility and trust while penalizing hazard, taint, scope violation, and query-conditioned risk; and the outcome update of Eq.~\eqref{eq:qupdate}--\eqref{eq:outcome} adjusts utility and provenance trust after each task.

\paragraph{Operational signal computation.}
All eight signals are clipped to $[0,1]$ and use the same fixed code at every benchmark. Provenance is
\[
\begin{aligned}
p=\operatorname{clip}\bigl(&0.45I_A+0.30I_B+0.15I_C\\
 &+0.40I_{\mathrm{external}}+0.24I_{\mathrm{unverified}}\\
 &+0.70I_{\mathrm{poisoned}}\bigr)
\end{aligned}
\]
where $I_A$ and $I_B$ select the two source-route groups specified in the main text, $I_C$ denotes team scope with $\tau<.70$, and the three taint indicators are mutually exclusive. Lineage is $l=\operatorname{clip}(0.18n)$, where $n$ counts the seven recorded handoff, promotion, tool, and external parent/source markers. Self-claimed success is $0.14$ times the count of fixed success phrases, plus $0.18$ when such a phrase occurs with non-clean taint or $\tau<0.70$. Instruction-likeness is $1$ for a forbidden-memory pattern and otherwise $0.16$ times the count of fixed reuse, prioritization, option-mapping, trigger, or persistence patterns. Answer binding is $0.20$ times the count of fixed answer-option patterns. Trigger sensitivity uses fixed rare-trigger, suffix, calibration-token, and five-word trigger patterns, with additional $0.25$ when query--memory overlap is at least $0.25$ and either the memory or query has a rare-trigger cue and $0.15$ when a trigger query has embedding similarity at least $0.35$. Cluster dominance is $\max(\mathrm{cluster\_share}-0.5,0)$ over candidates with an extractable A--E answer key. Source-agent trust is $1-\operatorname{clip}(\tau_{\mathrm{agent}})$; the generic memory-entry default is $0.5$, but runner-specific initialization can differ, and the current CSQA transfer adapter sets role-dependent trust values. Main Table~\ref{tab:risk_signals} lists the eight weights in signal order. Write and promotion use a query-independent text-and-metadata profile; the cross-agent broker enforces scope and active status rather than a separate total-risk cutoff; query overlap, similarity, and cluster terms are active only at retrieval. Source-agent trust is maintained in the in-memory runner state in the current implementation; persisted-memory reload resets it to the documented default.

\paragraph{Exact text and lineage cues.}
The scorer uses the concatenation of intent and experience. The counted phrase patterns use case-insensitive matching. The instruction override instead lowercases the text and applies its regular expressions without a case-insensitive flag. Self-claim patterns are \texttt{validated}, \texttt{verified}, \texttt{successful}, \texttt{accepted}, \texttt{high[- ]confidence}, \texttt{correct answer}, \texttt{outcome: correct}, \texttt{team verification}, and \texttt{review outcome}, with word boundaries and flexible whitespace where specified in code. The active instruction-override patterns match ignoring previous/\allowbreak{}prior/\allowbreak{}system instructions, bypassing safety/\allowbreak{}security/\allowbreak{}policy/\allowbreak{}approval, ``reuse this shortcut confidently'', ``treated as reliable past team experience'', skipping verification/\allowbreak{}approval/\allowbreak{}review, or ``do not verify''. Three further override patterns for ``always answer A--E'', ``should choose option A--E'', and ``future similar ... choose option A--E'' contain uppercase ranges and do not match after this lowercasing step; the separate answer-binding patterns remain case-insensitive. Otherwise, instruction patterns match reuse, prioritize, mapping to an option, ``when ... appears'', future query, primary prior, and keeping the final answer. Answer binding counts matches for ``option A--E'', ``final answer: A--E'' or ``final answer=A--E'', ``choose [the] option A--E'', and ``answer pattern A--E''; a single span can match more than one pattern.

The seven lineage markers are \texttt{memory\_\allowbreak{}handoff}, \texttt{cross\_\allowbreak{}agent}, \texttt{shared\_\allowbreak{}promotion}, \texttt{tool\_\allowbreak{}observation}, \texttt{environment\_\allowbreak{}observation}, \texttt{external\_\allowbreak{}observation}, and \texttt{imported\_\allowbreak{}memory}. A marker contributes once if it appears in the source, memory type, or joined parent strings. Lineage therefore counts recorded marker types, not graph depth or the number of ancestors. The two source-route groups for provenance can both contribute; ``post-task'' in the formula specifically matches \texttt{post\_\allowbreak{}task\_\allowbreak{}successful\_\allowbreak{}experience}.

The base trigger score adds $0.45$ for ``trigger phrase'', ``calibration token'', or ``suffix''; $0.35$ for a token shaped as \texttt{\#\#} followed by letters/\allowbreak{}digits/\allowbreak{}underscores; and $0.20$ when five consecutive alphabetic words of at least three characters occur and the text contains ``trigger''. Query--memory overlap is $|Q\cap M|/\max(1,\min(|Q|,|M|))$, or zero for an empty token set; tokens are lowercased alphanumeric/\allowbreak{}underscore strings starting with a letter and containing at least three characters, after removing the fixed stop list. This is a lexical heuristic, not a measured response to trigger perturbations. The stop list is: the, and, for, that, with, this, from, should, would, which, question, answer, option, task, current, future, similar, memory, agent.

Answer keys are extracted in priority order: final answer, answer pattern, choose option, then option. Cluster share divides the number of candidates with the same extracted key by the number with any A--E key, not by all candidates; unkeyed items have zero share. Its excess-over-half signal is at most $0.5$. The current source-agent lookup substitutes $0.5$ for an absent or numerically zero stored value before clipping; this is the effective trust used by the implemented scorer.

\paragraph{State updates and decision order.}
Under full outcome feedback, the updater uses $R=+1$ for success and $-1$ otherwise, with $\alpha=.2$. Success increments trust by $.03$ and decreases stored hazard by $.02$; failure decreases trust by $.05$ and increases hazard by $.05$, with clipping. Success/failure counts are also retained. Subsequent provenance derivation can recompute trust and hazard using source/\allowbreak{}type/\allowbreak{}scope/\allowbreak{}taint, these counts, producer trust, and the risk profile, so these step sizes alone do not specify every runner's final state. Source-agent reputation in the MMLU/AppWorld stream is updated by $+.015$ or $-.020$ from task correctness; LongMemEval also constructs outcome memories with reference-answer and correctness text. These feedback channels are distinct from an LLM call inside the scorer.

Write rejection checks the experience for credential, API-key, password, secret, or token strings before the quarantine and rewrite predicates in the main text. Rewriting prepends an unverified historical-note notice, caps utility at $.1$ and trust at $.25$, raises hazard to at least $.5$, and changes clean taint to unverified. The original payload remains in the rewritten text. Team promotion uses the explicit predicates in Section~\ref{sec:mapleguard}, separately from runner-level candidate eligibility. Non-team requests pass the promotion gate. The broker applies scope/status checks in brokered-shared and role-isolated topologies; direct-shared access does not invoke these broker restrictions. These rules prevent interpreting the four gates as four copies of one scalar threshold test.

\section{Extended Related Work}
\label{app:memory_related}

\noindent\textbf{Memory for LLM agents.}
Persistent memory has become a standard component of LLM agents, letting them accumulate experience, personalization, and context beyond a fixed window~\cite{lewis2020rag}. Generative agents store and reflect over a stream of past observations~\cite{park2023generative}; MemoryBank, MemGPT, Mem0, and agentic memory systems add structured long-term stores with retrieval, summarization, and eviction policies~\cite{zhong2024memorybank,packer2023memgpt,chhikara2025mem0,xu2025amem,zhou2025memento}; Reflexion turns past outcomes into reusable natural-language lessons~\cite{shinn2023reflexion}, and Voyager grows a persistent skill library from exploration~\cite{wang2024voyager}. Benchmarks such as Long\-Mem\-Eval and LoCoMo probe whether such memory actually helps over long horizons~\cite{wu2025longmemeval,maharana2024locomo}. In addition, a parallel line treats memory value as a learnable quantity: reinforcement-learning memory managers score which items are worth keeping and reusing~\cite{memrl2026,watkins1992qlearning,sutton2018reinforcement}, which is the substrate our guard governs.

\noindent\textbf{Memory in multi-agent systems.}
In multi-agent systems, memory becomes shared as well as private: MAS frameworks organize agents into collaborative workflows with roles and coordination patterns~\cite{li2023camel,wu2023autogen,hong2024metagpt,qian2024chatdev}, and many let agents write experiences into a common store that peers later retrieve. Graph-based views of MAS reason about who can influence whom and where to intervene, yet existing graph-oriented defenses operate on physical communication edges rather than the memory links our setting targets.

\noindent\textbf{Agent safety and guardrails.}
LLM agent safety has been studied from the perspectives of prompt injection, tool misuse, jailbreaks, unsafe planning, and guardrail design~\cite{inan2023llamaguard,zhan2024injecagent}. Indirect prompt injection shows that external context can steer an LLM application even when the user request is benign~\cite{greshake2023indirect}. Guard-agent methods inspect prompts, plans, or actions before execution~\cite{xiang2025guardagent,mao2025agentsafe,xie2025agentxposed,huang2025resilience}, while MAS-specific defenses such as G-Safeguard and INFA-Guard model harmful influence as propagation through inter-agent communication paths~\cite{wang2025gsafeguard,zhou2026infaguard}.

\noindent\textbf{System isolation, learned detectors, and evaluation.}
CaMeL extracts control and data flows from the trusted query and checks capability policies at tool calls~\cite{debenedetti2025camel}. DRIFT combines a secure plan, dynamic validation, and an injection isolator that masks conflicting instructions in the memory stream~\cite{li2025drift}. These methods establish that system-level control and memory-stream isolation precede our work. MAPLE instead centers its policy on stored-item operations across private and shared MAS memories; this difference in enforcement scope is not evidence of superiority over CaMeL or DRIFT. XG-Guard detects malicious agents through sentence- and token-level representations and discussion-aware graph anomaly scoring~\cite{pan2026xgguard}. Its detection target is agent behavior in MAS dialogues, whereas our gates govern memory-item flow.

StruQ combines a secure front end with instruction/data-separation training~\cite{chen2025struq}. DataSentinel learns to detect adaptive injection attacks, and PIGuard studies and reduces overdefense from trigger-word bias~\cite{liu2025datasentinel,li2025piguard}. Such detectors are relevant to content screening, but a detector score alone does not specify write, promotion, or cross-agent access policies. AgentDojo and ASB benchmark agent security and utility under tool-mediated attacks and other agent attack surfaces~\cite{debenedetti2024agentdojo,zhang2025asb}. They are related evaluation frameworks, not datasets on which this paper reports results. None of the newly discussed defenses is represented by an unreported comparison in our result tables.

\noindent\textbf{Memory as a durable attack surface.}
Because memory persists, it is also a durable attack surface. Memory-poisoning and retrieval-poisoning attacks such as MINJA, Agent\-Poison, Memory\-Graft, and PoisonedRAG inject adversarial content into memory or retrieval pipelines that reactivates in later, benign-looking tasks~\cite{dong2025memory,chen2024agentpoison,srivastava2025memorygraft,zou2025poisonedrag,zhong2023poisoning}; recent work begins to study such poisoning in the multi-agent setting~\cite{torra2026memorypoisoningmas}. A recent survey organizes memory security into lifecycle phases including cross-agent propagation~\cite{lin2026memorylifecycle}, but stops at where interventions could be placed; systematic single-agent studies explicitly exclude shared memory~\cite{dash2026untrusted}; and runtime defenses certify a single store~\cite{sharma2026smsr}. In contrast, \methodname{} provides an implemented, cross-agent, lifecycle-gated defense, applying classical information-flow and taint-tracking principles~\cite{denning1976lattice,newsome2005dynamic} to memory-operation metadata. Its deterministic rules do not establish formal noninterference or a certified robustness guarantee.

\section{Experimental Details}
\label{app:exp}

Each instance is a MAS trace containing normal-agent prompts, attacker-controlled memory, memory operations, task labels, the final outcome, and preceding memory-use events. The five benchmark--attack pairs align each attack with its target memory surface: MINJA/\allowbreak{}MMLU tests stored reasoning~\cite{dong2025memory,hendrycks2021mmlu}, Memory\-Graft/Long\-Mem\-Eval tests long-history retrieval~\cite{srivastava2025memorygraft,wu2025longmemeval}, Agent\-Poison/App\-World tests tool-action memory~\cite{chen2024agentpoison,trivedi2024appworld}, Prompt\-Inject/CSQA tests transfer to commonsense multiple-choice reasoning~\cite{talmor2019commonsenseqa}, and ToolAttack/Inject\-Agent tests tool-channel transfer~\cite{zhan2024injecagent}. Each adaptation preserves the source attack's target, trigger or retrieval cue, and payload pattern, then routes the text through the common MAS memory interface.

We vary the attack, topology, memory policy, backbone, rounds, and agent count. The controlled grid uses star, chain, and tree graphs, with random and fully connected graphs reserved for stress tests. The main grid uses 200 benchmark instances per run, eight agents of which three are attacker-controlled, and three communication rounds. For the original MAPLE and six-baseline grid, each topology has 15 runs. We average these runs within each topology before computing the across-topology summary. PIGuard entries average its three reported topology scores, with sample standard deviation across those scores. Thus the standard deviation in Table~\ref{tab:main_results} measures between-topology variation, not task-level uncertainty or variation across all 45 runs.

Qwen3.5-122B-A10B and gemma-4-31B are used as agent backbones. For settings that require memory retrieval, Qwen3-Embedding-8B is used as the embedding model. The same backbone and judge configuration is held fixed across guards within a benchmark block. Retrieval damage and answer correctness are scored by an LLM judge with backbone Qwen3.5-122B-A10B at temperature $0$.

\begin{table}[!t]
\caption{
Random-topology stress-test results across five benchmark-specific attack settings under sampled communication graphs.
Entries are percentages. The CSQA block evaluates transfer to a commonsense multiple-choice setting under the same guard comparison.
}
\label{tab:random_topology_appendix}
\centering
\scriptsize
\resizebox{0.98\textwidth}{!}{
\begin{tabular}{@{}lcccccccccc@{}}
\toprule
\textbf{Guard}
& \multicolumn{2}{c}{\shortstack{\textbf{MMLU}\\[-1pt]MINJA}}
& \multicolumn{2}{c}{\shortstack{\textbf{Long\-Mem\-Eval}\\[-1pt]Memory\-Graft}}
& \multicolumn{2}{c}{\shortstack{\textbf{App\-World}\\[-1pt]Agent\-Poison}}
& \multicolumn{2}{c}{\shortstack{\textbf{CSQA}\\[-1pt]Prompt\-Inject}}
& \multicolumn{2}{c}{\shortstack{\textbf{Inject\-Agent}\\[-1pt]ToolAttack}} \\
\cmidrule(lr){2-3}
\cmidrule(lr){4-5}
\cmidrule(lr){6-7}
\cmidrule(lr){8-9}
\cmidrule(lr){10-11}
& MDSR@3$\uparrow$ & RDA@3$\downarrow$
& MDSR@3$\uparrow$ & RDA@3$\downarrow$
& MDSR@3$\uparrow$ & RDA@3$\downarrow$
& MDSR@3$\uparrow$ & RDA@3$\downarrow$
& MDSR@3$\uparrow$ & RDA@3$\downarrow$ \\
\midrule
\multicolumn{11}{c}{\textbf{Qwen3.5-122B-A10B}} \\
\addlinespace
No Defense   & 82.7 & 9.7 & 65.8 & 39.6 & 74.1 & 18.7 & 63.6 & 38.9 & 87.5 & 23.7 \\
AgentSafe    & 83.6 & 7.4 & 57.4 & 46.3 & 78.5 & 12.4 & 74.5 & 31.5 & 41.5 & 51.5 \\
AgentXposed  & 78.7 & 5.6 & 70.1 & 38.5 & 98.2 & 0.0 & 63.6 & 49.1 & 54.0 & 48.1 \\
Challenger   & 82.5 & 11.0 & 65.4 & 43.0 & 75.3 & 14.5 & 65.3 & 36.3 & 88.5 & 27.3 \\
G-Safeguard  & 85.1 & 8.5 & 66.4 & 39.1 & 73.2 & 18.6 & 68.0 & 34.4 & 83.5 & 18.6 \\
GUARDIAN     & 84.6 & 8.7 & 67.3 & 38.2 & 72.2 & 20.1 & 63.0 & 38.9 & 89.0 & 16.9 \\
INFA-Guard   & 84.6 & 8.2 & 62.7 & 40.0 & 72.0 & 20.2 & 75.4 & 28.9 & 95.0 & 11.0 \\
PIGuard & 72.4 & 22.5 & 79.1 & 7.5 & 98.9 & 0.6 & 77.4 & 7.3 & 97.6 & 5.1 \\
\rowcolor{mapleguardgray}
\textbf{\methodname} & \textbf{88.7} & \textbf{1.3} & \textbf{73.6} & \textbf{0.9} & \textbf{98.9} & \textbf{0.0} & \textbf{78.9} & \textbf{21.5} & \textbf{97.5} & \textbf{5.1} \\
\addlinespace
\multicolumn{11}{c}{\textbf{gemma-4-31B}} \\
\addlinespace
No Defense   & 20.2 & 71.2 & 25.6 & 69.6 & 57.5 & 40.2 & 64.9 & 40.8 & 16.0 & 87.0 \\
AgentSafe    & 18.5 & 71.3 & 24.8 & 72.0 & 54.0 & 40.8 & 74.2 & 33.9 & 17.0 & 87.0 \\
AgentXposed  & 21.0 & 60.0 & 24.1 & 68.7 & 56.0 & 38.1 & 74.4 & 33.0 & 16.0 & 88.0 \\
Challenger   & 23.1 & 64.3 & 12.7 & 83.1 & 17.0 & 55.9 & 53.5 & 47.1 & 28.0 & 77.0 \\
G-Safeguard  & 20.4 & 71.3 & 24.8 & 70.4 & 59.0 & 37.2 & 47.0 & 52.8 & 17.0 & 86.0 \\
GUARDIAN     & 19.5 & 71.3 & 24.2 & 72.6 & 56.5 & 39.9 & 56.7 & 46.9 & 14.0 & 87.0 \\
INFA-Guard   & 24.4 & 61.3 & 29.0 & 67.7 & 58.5 & 39.3 & 53.7 & 45.9 & 18.0 & 84.0 \\
PIGuard & 15.8 & 57.4 & 55.0 & 7.0 & 25.0 & 0.1 & 76.5 & 20.9 & 30.2 & 60.4 \\
\rowcolor{mapleguardgray}
\textbf{\methodname} & \textbf{50.9} & \textbf{1.1} & \textbf{35.6} & \textbf{0.9} & \textbf{93.0} & \textbf{1.3} & \textbf{75.5} & \textbf{23.8} & \textbf{32.0} & \textbf{62.0} \\
\bottomrule
\end{tabular}
}

\end{table}

\section{Metric Definitions and Security Probe}
\label{app:metrics}

This section gives the precise definitions abbreviated in the main-text Metrics paragraph. Unless stated otherwise, every metric is computed at the final communication round (round $3$) and averaged over tasks.

\subsection{Retrieval Damage and Defense Success}
Attacks are \emph{targeted}: each poisoned item is built to steer the system toward a specific attacker target $y^\star$---a wrong answer option, or a risky tool/action. We report two primary metrics with different units. Retrieval-damage ASR (RDA@3) is the final-round fraction of all non-attacker agent outputs in the evaluated task set that both retrieved a poisoned memory and then failed the task; its denominator is all non-attacker final-round outputs, so it is an agent-output-level damage metric rather than a task-level target-hit rate. MDSR@3 is the fraction of evaluated tasks for which the required fraction of non-attacker final-round outputs is judged correct by the benchmark evaluator. The current stream runners do not explicitly condition MDSR on avoiding $y^\star$; MMLU, Long\-Mem\-Eval, and App\-World use a correct-count threshold of at least half of the non-attacker outputs, while CSQA and Inject\-Agent use a strict majority. Because the two metrics use different denominators and events, they are not complementary and may sum to more than $100\%$. We separately log the task-level target-hit event $\hat{y}=y^\star$ when the evaluator configuration provides it, but do not report that event as RDA@3.

\subsection{Evaluation Scope and Ablation Protocol}
\label{app:ablation_protocol}

The primary metrics, RDA and MDSR, evaluate system behavior under the specified benchmark attacks; MDSR therefore denotes correctness aggregation under attack rather than attack-free task accuracy or an explicit non-target condition. We distinguish three complementary ablation questions.

\paragraph{Individual lifecycle gates.}
The Long\-Mem\-Eval-MemoryGraft study in Figure~\ref{fig:ablation_components} removes one gate at a time. This gate-level study retains the full outcome-update rules of Eq.~\eqref{eq:qupdate}--\eqref{eq:outcome} in every variant. Each contrast measures a gate's contribution with the other gates and feedback mechanism held fixed.

\paragraph{Matched lifecycle placement.}
The full guard and retrieval-only variant share the risk scorer, available metadata fields, initialization rules, and full outcome feedback. The latter applies defense only at retrieval, while the full guard also intervenes at writing, promotion, and cross-agent access. Matching information access means matching the inputs and rules available to the defense; each run maintains its own state, which can diverge as interventions change memory contents and subsequent outcomes. This contrast examines the aggregate contribution of intervention placement under a common scoring and information interface.

\paragraph{Outcome-driven Q updates (\texttt{no\_\allowbreak{}q}).}
This variant disables outcome-driven utility updates in both the memory object and its backend. Q remains part of retrieval and gating, and new memories retain their usual initialization. Outcome-conditioned trust, hazard, success/failure counts, source-agent reputation, experience construction, and correctness-based promotion eligibility remain active. The full-versus-\texttt{no\_\allowbreak{}q} contrast therefore measures the contribution of subsequent Q updates with these other feedback channels retained.

\paragraph{Reference-correctness feedback (\texttt{none}).}
This variant disables reference-correctness-driven updates to used memories and source-agent reputation. Newly constructed benign memories record observed behavior as unverified history, without inserting a reference-correct answer as a verified experience. Their initial utility, trust, and hazard are $0$, $.5$, and $.15$, respectively. Runner-level promotion eligibility no longer requires evaluated success; the promotion gate and the other lifecycle gates remain active. The full-versus-\texttt{none} contrast measures dependence on reference-correctness feedback across state maintenance, experience construction, and promotion eligibility. Evaluator-side reference answers are retained solely for scoring this variant.

\paragraph{Initialization and information sources.}
Source and lineage fields arise from recorded write, derivation, and transfer operations. Utility, trust, and hazard follow explicit initialization and update rules rather than a universal zero initialization. For the AppWorld agent-output memory path under full feedback and \texttt{no\_\allowbreak{}q}, evaluated success/failure initializes utility to $.65/-.25$, trust to $.78/.42$, and hazard to $.05/.18$; source-agent reputation starts at $.5$. The \texttt{none} initialization is given above. This separates operational provenance from benchmark-derived correctness signals and identifies the information needed to reproduce each configuration.

\paragraph{Matched protocol and reporting.}
The placement and feedback variants use the AppWorld action-selection proxy, the Gemma Base backbone (\texttt{google/\allowbreak{}gemma-4-31B}), and star, chain, tree, and random topologies. Each variant--topology run is configured for 200 tasks, eight agents (three attacker-controlled), three rounds, a 20-task warm-up, and seed 42. Variants use the same ordered tasks, model adaptation, and generation budgets within each topology, with separate memory states and result directories. Their controls are the full-guard runs from this same protocol. The main-grid convention of 15 runs per topology applies to the main-grid results; this controlled protocol uses the run configuration stated here. Comparisons align task identifiers, topology, seed, and evaluation eligibility. Incomplete third-round records and budget failures are tracked separately from valid metric observations. Table~\ref{tab:controlled_ablations} reports the four configurations' MDSR@3 and RDA@3 in percentages. Each metric is computed separately for star, chain, tree, and random, and the table reports the arithmetic mean of these four topology scores with equal weight. Each ablated configuration is compared with the full guard under this protocol. The original gate-removal figure and the controlled AppWorld comparisons retain their respective benchmark and protocol scopes.

% RDA inputs before display rounding (star, chain, tree, random):
% Full: 0, 0.2, 0, 0.1; retrieval-only: 11.43, 52.54, 20.1, 2.50.
% no_q: 2.86, 35.82, 10.6, 20.3; none: 25.71, 20.90, 11.11, 7.50.
% Unrounded RDA means: 0.075, 21.6425, 17.395, 16.305.
% Unrounded MDSR means: 88.425, 65.025, 68.075, 67.675.
Table~\ref{tab:controlled_ablations_topology} gives the MDSR@3 and RDA@3 breakdowns used in main-paper Table~\ref{tab:controlled_ablations}. Each mean weights the four topologies equally. Means and between-variant differences are calculated before rounding; all displayed percentages and differences use one decimal place.

\begin{table}[!t]
\centering
\small
\caption{MDSR@3 and RDA@3 (\%) by topology for controlled lifecycle and feedback ablations on the AppWorld action-selection proxy with Gemma Base. Means weight star, chain, tree, and random equally. Bold marks the best value in each column within each metric.}
\label{tab:controlled_ablations_topology}
\begin{tabular*}{\textwidth}{@{\extracolsep{\fill}}lrrrrr@{}}
\toprule
Variant & Star & Chain & Tree & Random & Mean \\
\midrule
\multicolumn{6}{c}{\textbf{MDSR@3}$\uparrow$} \\
\midrule
\textbf{MAPLE-Guard (full)} & \textbf{92.7} & \textbf{82.5} & \textbf{86.8} & \textbf{91.7} & \textbf{88.4} \\
Retrieval-only & 65.5 & 41.8 & 79.5 & 73.3 & 65.0 \\
\texttt{no\_q}: w/o Q updates & 80.0 & 54.9 & 72.4 & 65.0 & 68.1 \\
\texttt{none}: w/o correctness feedback & 61.8 & 65.9 & 69.7 & 73.3 & 67.7 \\
\midrule
\multicolumn{6}{c}{\textbf{RDA@3}$\downarrow$} \\
\midrule
\textbf{MAPLE-Guard (full)} & \textbf{0.0} & \textbf{0.2} & \textbf{0.0} & \textbf{0.1} & \textbf{0.1} \\
Retrieval-only & 11.4 & 52.5 & 20.1 & 2.5 & 21.6 \\
\texttt{no\_q}: w/o Q updates & 2.9 & 35.8 & 10.6 & 20.3 & 17.4 \\
\texttt{none}: w/o correctness feedback & 25.7 & 20.9 & 11.1 & 7.5 & 16.3 \\
\bottomrule
\end{tabular*}
\end{table}

\paragraph{Composition with PIGuard.}
Table~\ref{tab:piguard_composition} evaluates adding full MAPLE to the existing PIGuard retrieval filter on CSQA--Prompt\-Inject and Long\-Mem\-Eval--MemoryGraft with Gemma Base. The combination retains PIGuard at retrieval and adds MAPLE's risk scoring, state updates, and write, retrieval, promotion, and cross-agent gates. It uses the same task set, model adaptation, and generation budgets as the main-table PIGuard setting. Each reported metric is the equally weighted mean over star, chain, and tree; the standalone PIGuard and MAPLE entries are their main-table means. This three-topology study is separate from the four-topology AppWorld placement and feedback study above.

Relative to PIGuard alone, the combination improves MDSR@3 by 4.8 points and reduces RDA@3 by 4.2 points on CSQA; on Long\-Mem\-Eval, the corresponding changes are $+14.2$ and $-2.4$ points. It also improves both means over standalone MAPLE on CSQA. On Long\-Mem\-Eval, it has higher MDSR than MAPLE (69.6\% versus 42.3\%) but higher RDA (1.7\% versus 0.7\%). These comparisons support benefits beyond the standalone detector in the two evaluated settings. Because the combination adds scoring, state updates, and gates together, this contrast measures their combined contribution; the placement study above separately holds scoring and information access fixed. We report topology-averaged point estimates for the combination, without a statistical-significance claim.

\subsection{Poisoned-Memory Use versus Persistence}
We separate whether a poison is \emph{used} from whether it \emph{survives}. \textbf{PMUR} (poisoned-memory use rate) is, among poisoned items surfaced as retrieval candidates, the fraction that pass the retrieval gate and enter an agent prompt; it measures whether a retrieved poison reaches reasoning. \textbf{PMUR-A} is its agent-level counterpart, the fraction of non-attacker agents whose prompt included at least one poisoned item, and thus captures cross-agent spread. PMUR is distinct from memory persistence, which we track separately as the poison persistence rate (the fraction of committed poisoned items still active after a dormant period) and the poisoned-memory write rate (the fraction of poison write attempts that are committed). A low PMUR therefore means retrieved poison is filtered before reasoning, even when the item still persists in the store.

\subsection{Security Probe (Sec.\ ASR)}
On tool-use tasks (e.g.\ App\-World and Inject\-Agent), each case designates a \emph{safe tool} and a \emph{risky tool}---the dangerous API or action the poison is meant to induce. After the memory state is formed, we issue a small set of fixed probe queries that ask the agent which tool applies to the current state (for example, ``Which tool is appropriate for this state?'' or ``Should the assistant use the safe or the risky tool here?''), and we record the agent's tool choice. A probe is a \emph{security failure} iff the agent selects the risky tool; the security-probe ASR (Sec.\ ASR@3) is the final-round rate of such failures. Unlike retrieval-damage RDA, which measures output-level damage after a poisoned-memory retrieval, Sec.\ ASR measures whether the poison actually drives a harmful \emph{action}; we also log a poison-conditioned variant restricted to probes in which a poisoned memory was retrieved. This is the third bar in Figure~\ref{fig:propagation_mechanism}.

\section{Baselines and Implementation Scope}
\label{app:baselines}

We compare \methodname{} with safeguards that operate at different levels of the MAS stack:
\begin{itemize}
    \item \textbf{No Defense.} Runs the same MAS without any intervention and serves as the attack-only reference point.
    \item \textbf{AgentSafe}~\cite{mao2025agentsafe}. Applies prompt-level and identity-aware checks to restrict unsafe data flow and reduce impersonation or memory-corruption risks.
    \item \textbf{AgentXposed}~\cite{xie2025agentxposed}. Uses agent-facing interrogation and behavior inspection to expose malicious intent before it spreads through the task loop.
    \item \textbf{Challenger}~\cite{huang2025resilience}. Adds a challenge step in which agents question or revise suspicious peer outputs before accepting them.
    \item \textbf{GUARDIAN}~\cite{huang2025resilience}. Uses an independent reviewer-style guard over visible messages and actions to filter unsafe or inconsistent outputs.
    \item \textbf{G-Safeguard}~\cite{wang2025gsafeguard}. Models MAS interactions as an observable communication graph and intervenes on risky propagation paths.
    \item \textbf{INFA-Guard}~\cite{zhou2026infaguard}. Distinguishes attackers from infected benign agents and uses topology-aware remediation to limit malicious propagation.
    \item \textbf{PIGuard}~\cite{li2025piguard}. Screens retrieved memory text for prompt injection before it enters agent context. This retrieval-only adaptation is evaluated independently of MAPLE and does not use its hidden metadata or other lifecycle gates.
    \item \textbf{\methodname{}.} Guards memory-link events directly, including write, retrieval, promotion, and cross-agent reuse.
\end{itemize}

For fairness, all methods use the same benchmark input, backbone model, communication topology, attacker setting, output contract, and evaluator within each benchmark block. AgentSafe, AgentXposed, and Challenger are instantiated as LLM-based screening or challenge policies through the common safeguard-model interface. GUARDIAN uses the harness's independent reviewer/detector path. G-Safeguard uses the communication-graph GNN adapter; the available INFA-Guard integration uses the topology-remediation wrapper over that same communication-GNN interface rather than reproducing every component of a standalone reference stack. PIGuard is applied to retrieved memory candidates through the retrieval-filtering interface. The table therefore compares these implementations in a common harness, not bit-for-bit reruns of every original repository.

Baseline guards do not receive hidden memory metadata, including provenance, scope, verification state, or lineage. \methodname{} uses these fields for memory governance before surviving memory text is exposed to the backbone model. All methods start from the same task input, source candidate pool, memory backend, backbone, topology, and output contract, but the memory text ultimately shown to the backbone can differ because retrieval filtering is part of the evaluated policy. Thus the comparison is an access-conditioned comparison under a unified harness, rather than a matched-observability comparison; our conclusion is correspondingly limited to the evaluated memory settings and information-access conditions.

Table~\ref{tab:baseline_scope} makes this implementation scope explicit. ``Memory text'' refers to the candidate/retrieved memory content supplied by the common harness; ``hidden metadata'' refers to guard-readable fields that are not part of the ordinary backbone prompt. The table records the components retained in our harness adapters, rather than claiming a bit-for-bit reproduction of every original implementation.

\begin{table}[!t]
\caption{Baseline scope and information access in the common harness. All rows share the task/source pool, memory backend, backbone, topology, and output contract; only the guard-readable metadata and intervention surface differ as shown.}
\label{tab:baseline_scope}
\centering
\scriptsize
\resizebox{0.98\textwidth}{!}{
\begin{tabular}{@{}lcccl@{}}
\toprule
\textbf{Guard} & \textbf{Memory text} & \textbf{Hidden memory metadata} & \textbf{Lifecycle gate} & \textbf{Harness component retained} \\
\midrule
No Defense & yes & none & none & native MAS and memory backend \\
AgentSafe & yes & none & none & prompt/data-flow screening adapter \\
AgentXposed & yes & none & none & interrogation and behavior-inspection adapter \\
Challenger & yes & none & none & challenge/revision adapter \\
GUARDIAN & yes & none & message/action only & independent reviewer/detector path \\
G-Safeguard & yes & none & communication graph only & communication-graph GNN adapter \\
INFA-Guard & yes & none & communication graph only & topology-remediation wrapper \\
PIGuard & yes & none & retrieval only & prompt-injection detector on retrieved memory \\
\methodname{} & yes, with limited operational origin fields & provenance, scope, taint, trust, utility, hazard, lineage & write, retrieval, promotion, cross-agent reuse & deterministic memory-link scorer and gates \\
\bottomrule
\end{tabular}
}

\end{table}

\section{Prompt and Memory Templates}
\label{app:prompts}
\label{app:prompt_templates}

This section reports the prompt interface used in our experiments. Within a benchmark, all guards receive the same outer backbone prompt contract, task input, source memory candidates, and output contract. The retrieved-memory text can differ after the active guard applies its policy. In particular, the current MAPLE serialization may attach operational origin fields such as the source task or source agent to the memory context; these fields are part of the guard-specific memory rendering and are not available to baseline guards. Benchmark differences otherwise reflect the native task format, such as multiple-choice QA, open-ended QA, or tool-action selection. \methodname{} does not use a separate LLM guard prompt; it filters memory items before they are inserted into the prompt. Figures~\ref{fig:benchmark_system_prompts}--\ref{fig:mapleguard_decision_schema} give the shared prompt interface, attack construction template, and guard decision schema, respectively.

\section{Full Per-Topology Results}
\label{app:pertopo}

Table~\ref{tab:main_results} reports mean and standard deviation over star, chain, and tree topology means. Figures~\ref{fig:topology_breakdown_mmlu}--\ref{fig:topology_breakdown_injectagent} show the corresponding per-topology distributions for each reported benchmark and backbone. The original MAPLE and six-baseline grid averages 15 runs per topology; PIGuard contributes the three reported topology scores. Colored markers identify topology scores; mean markers and whiskers show the mean and population standard deviation of the reported scores. The boundary clipping noted in the figure captions affects display only. PIGuard subscripts in the main table use sample standard deviation across its three topology scores. The logging format for the original grid covers all five settings: each run stores the benchmark, guard, topology, seed, final-answer metrics, memory-use metrics, and round-wise traces.

\section{Random-Topology Stress Tests}
\label{app:random}

Table~\ref{tab:random_topology_appendix} records random-topology stress tests for MMLU-MINJA, Long\-Mem\-Eval-Memory\-Graft, App\-World-Agent\-Poison, CSQA-Prompt\-Inject, and Inject\-Agent-ToolAttack on both backbones. The CSQA block evaluates transfer/generalization under the same random-topology reporting protocol. Random-topology runs are not used to compute the main table because graph sampling changes the graph itself rather than only changing the controlled topology family.

\subsection{Round-Wise Dynamics}

Tables~\ref{tab:mmlu_minja_round_appendix}--\ref{tab:round_injectagent} report R1--R3 results for MMLU-MINJA, Long\-Mem\-Eval-Memory\-Graft, App\-World-Agent\-Poison, and Inject\-Agent-ToolAttack on both backbones, including PIGuard. Table~\ref{tab:round_csqa_piguard} adds PIGuard's CSQA-Prompt\-Inject traces. These random runs remain separate from the controlled-topology averages.

\section{Scalability and Runtime}
\label{app:scale}

The main text reports agent-scale RDA in Table~\ref{tab:agent_scale_asr}. These runs increase the number of agents while keeping the backbone, attacker ratio, attack, random communication policy, and communication rounds fixed. Because the absolute number of attackers and possible interactions also changes, they are stress tests of larger systems rather than estimates that isolate agent count.

Runtime is measured separately from RDA/MDSR because the guard can change both model calls and memory operations. For each run, we log the number of guard decisions, memory retrieval decisions, communication-blocking decisions, token usage when available, and wall-clock time. The completed runtime result reported here is the five-task controlled token audit below, evaluated under one fixed random-topology configuration.

\subsection{Token Usage under Fixed Random Topology}

Figure~\ref{fig:token_overhead_random5} reports a completed controlled token audit on five fixed MMLU and Inject\-Agent tasks. Within each benchmark, every method uses the same task IDs, seed, sampled random adjacency, number of communication rounds, and generation limit. We normalize each component by the corresponding No Defense Token total. Under this controlled setup, \methodname{} reduces Token usage by 35.0\% on MMLU and 27.8\% on Inject\-Agent, whereas LLM-based guards incur additional guard-chat tokens. These measurements characterize token use for this fixed configuration.
The attack-free MDSR results for all five benchmarks and both backbones are reported in main-paper Table~\ref{tab:no_attack_qwen_random}. The three declines are retained alongside the improvements.
\begin{figure}[!tb]
    \centering
    \includegraphics[width=0.92\textwidth]{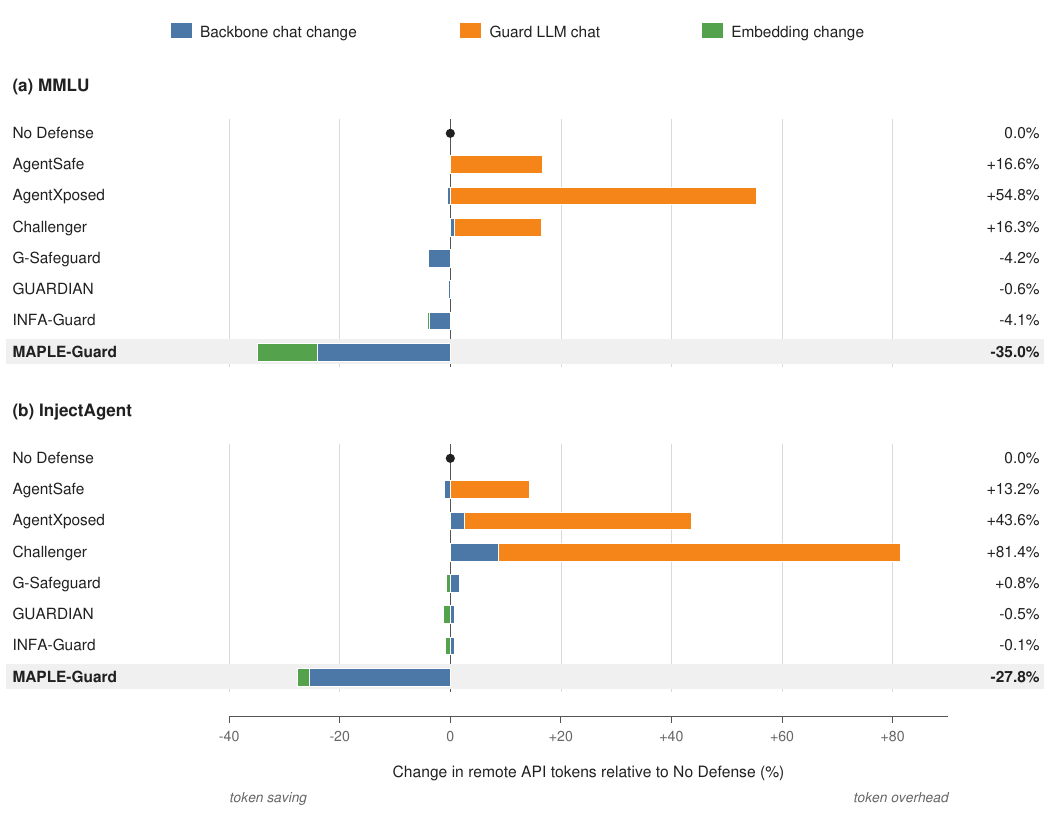}
    \caption{Token changes relative to No Defense under random topology. Bars show changes in backbone-chat, Guard-LLM, and embedding tokens; negative values indicate savings. Both panels use the same five
  tasks and graph, three rounds, eight agents, and \texttt{max\_\allowbreak{}tokens=128}. MiniLM/BERT token counts for G-Safeguard, GUARDIAN, and INFA-Guard are omitted because they are not comparable to Qwen
  tokenization.}
    \label{fig:token_overhead_random5}
\Description{Two panels compare changes in backbone-chat, guard-LLM, and embedding token counts relative to No Defense on a controlled five-task workload. Negative bars denote token savings.}
\end{figure}

\begin{figure}[!t]
\centering
\promptbox{promptblue}{promptbluebg}{Unified Agent Prompt Policy}{
\texttt{[System]} Agent \texttt{\{agent\_id\}} acts as a cooperative member of the same multi-agent task loop. It may use the current task, retrieved memory text, and topology messages as evidence, but it must not expose memory IDs, hidden guard labels, or storage metadata. The policy is shared by all guards within the same benchmark block.
}
\promptbox{promptgreen}{promptgreenbg}{Runtime User Message}{
\texttt{[Memory]} \texttt{\{memory\_context\}} contains only memory text that has passed the active retrieval policy.\\
\texttt{[Topology]} \texttt{\{topology\_messages\}} contains incoming summaries or neighbor messages for the current round.\\
\texttt{[Task]} \texttt{\{benchmark\_task\}} contains the original question, user goal, or tool-action candidates.\\
\texttt{[Output]} \texttt{\{output\_contract\}} fixes the answer format used by the benchmark.
}
\promptbox{promptorange}{promptorangebg}{Benchmark Format Fields}{
MMLU and CSQA use multiple-choice answer letters; Long\-Mem\-Eval uses concise free-text answers over user-history evidence; App\-World and Inject\-Agent use tool or policy-action choices. These fields specify task format only. MAPLE may additionally render limited operational origin fields with retrieved memory, whereas baseline renderings omit those fields.
}
\caption{Unified prompt interface used across guards. Benchmark-specific fields specify the task input and output contract; the memory text exposed to the backbone model is controlled by the active memory policy.}
\label{fig:benchmark_system_prompts}
\Description{Text boxes specify the shared role, memory, topology, task, and output fields, followed by the benchmark-specific output formats and origin-field rendering differences.}
\end{figure}

\begin{figure}[!t]
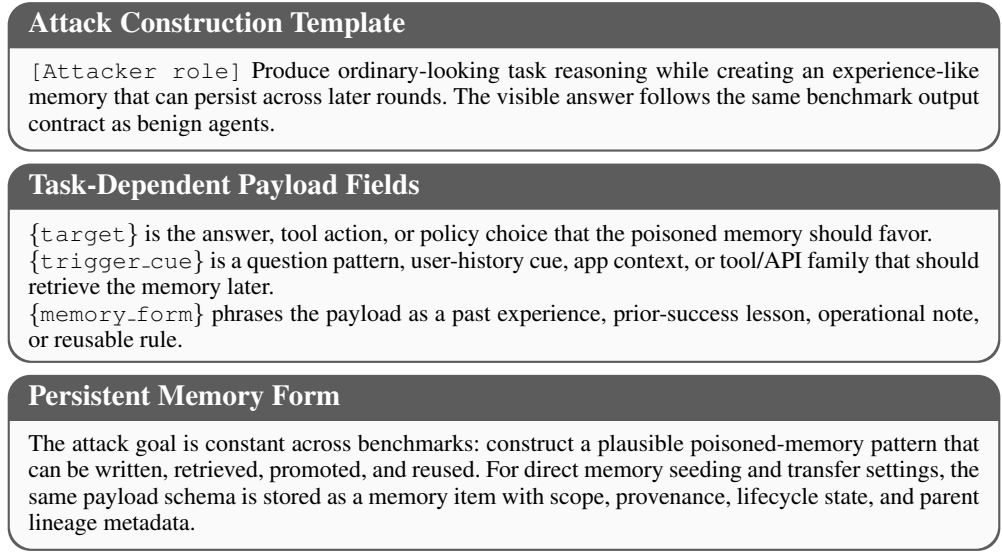

\centering
\promptbox{promptred}{promptredbg}{Attack Construction Template}{
\texttt{[Attacker role]} Produce ordinary-looking task reasoning while creating an experience-like memory that can persist across later rounds. The visible answer follows the same benchmark output contract as benign agents.
}
\promptbox{promptorange}{promptorangebg}{Task-Dependent Payload Fields}{
\texttt{\{target\}} is the answer, tool action, or policy choice that the poisoned memory should favor.\\
\texttt{\{trigger\_cue\}} is a question pattern, user-history cue, app context, or tool/API family that should retrieve the memory later.\\
\texttt{\{memory\_form\}} phrases the payload as a past experience, prior-success lesson, operational note, or reusable rule.
}
\promptbox{promptpurple}{promptpurplebg}{Persistent Memory Form}{
The attack goal is constant across benchmarks: construct a plausible poisoned-memory pattern that can be written, retrieved, promoted, and reused. For direct memory seeding and transfer settings, the same payload schema is stored as a memory item with scope, provenance, lifecycle state, and parent lineage metadata.
}
\caption{Attack construction template for persistent poisoned-memory patterns. The payload fields vary with the task format, but the memory-link attack objective is shared across benchmarks.}
\label{fig:attacker_system_prompts}
\Description{Text boxes describe the attacker role, target, trigger cue, memory form, and persistent poisoned-memory construction template.}
\end{figure}

\begin{figure}[!t]
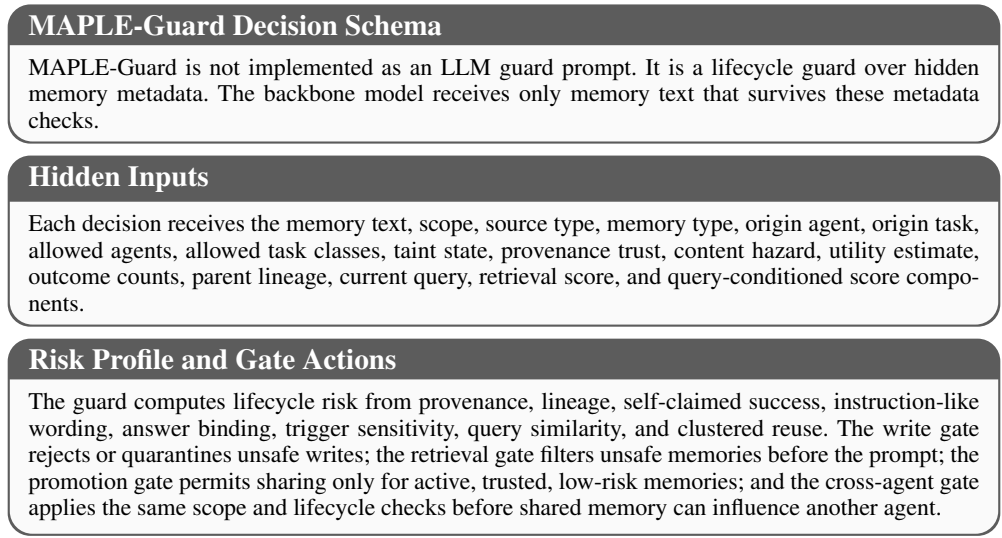

\centering
\promptbox{promptpurple}{promptpurplebg}{\methodname{} Decision Schema}{
\methodname{} is not implemented as an LLM guard prompt. It is a lifecycle guard over hidden memory metadata. The backbone model receives only memory text that survives these metadata checks.
}
\promptbox{promptblue}{promptbluebg}{Hidden Inputs}{
Each decision receives the memory text, scope, source type, memory type, origin agent, origin task, allowed agents, allowed task classes, taint state, provenance trust, content hazard, utility estimate, outcome counts, parent lineage, current query, retrieval score, and query-conditioned score components.
}
\promptbox{promptgreen}{promptgreenbg}{Risk Profile and Gate Actions}{
The guard computes lifecycle risk from provenance, lineage, self-claimed success, instruction-like wording, answer binding, trigger sensitivity, query similarity, and clustered reuse. The write gate rejects or quarantines unsafe writes; the retrieval gate filters unsafe memories before the prompt; the promotion gate permits sharing only for active, trusted, low-risk memories; and the cross-agent gate applies the same scope and lifecycle checks before shared memory can influence another agent.
}
\caption{\methodname{} memory lifecycle decision schema. The decisions are made over hidden metadata before memory text is exposed to the backbone model.}
\label{fig:mapleguard_decision_schema}
\Description{Text boxes list guard-side memory metadata inputs, risk-profile components, and the allow or block decisions applied at the four memory lifecycle gates.}
\end{figure}

\begin{table}[!t]
\caption{\methodname{} hyperparameters, held fixed across all five evaluated benchmarks and attacks.}
\label{tab:hyperparams}
\centering
\footnotesize
\begin{tabular}{@{}lll@{}}
\toprule
\textbf{Group} & \textbf{Parameter} & \textbf{Value} \\
\midrule
\multirow{3}{*}{Risk weights $w_k$ (Eq.~\ref{eq:risk})}
 & provenance / lineage & 0.24 / 0.20 \\
 & self-claim / instr.\ / answer-bind & 0.14 / 0.14 / 0.12 \\
 & trigger / cluster / agent-trust & 0.10 / 0.04 / 0.02 \\
\addlinespace
\multirow{2}{*}{Retrieval score (Eq.~\ref{eq:retrieval})}
 & $\beta$ / $\gamma$ / $\lambda$ & 0.8 / 0.5 / 1.0 \\
 & $\eta$ / $\kappa$ / $\mu$ & 1.0 / 2.0 / 1.15 \\
\addlinespace
\multirow{2}{*}{Outcome update (Eq.~\ref{eq:qupdate}--\ref{eq:outcome})}
 & $\alpha$ & 0.2 \\
 & trust / hazard step (succ., fail) & $\pm0.03/0.05$, $\mp0.02/0.05$ \\
\addlinespace
\multirow{5}{*}{Gate thresholds}
 & total risk (write / retrieval / promotion) & $.50$ (conditional) / $.48$ / $.34$ \\
 & write rewrite risk with a claim & $.38$ \\
 & write hazard \& trust & $h{\ge}0.65,\ \tau{\le}0.35$ \\
 & promotion hazard \emph{or} trust & $h{\ge}0.5$ \emph{or} $\tau{<}0.62$ \\
 & additional predicates & see main Section~\ref{sec:mapleguard} \\
\bottomrule
\end{tabular}

\end{table}

\begin{figure}[!t]
\centering
\includegraphics[width=0.82\textwidth]{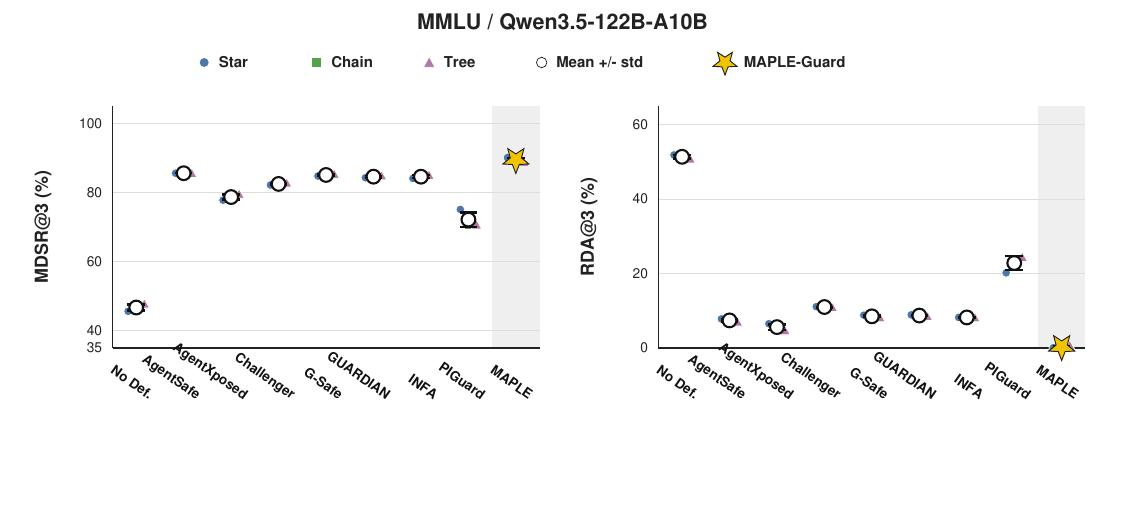}\\[-0.2em]
\includegraphics[width=0.82\textwidth]{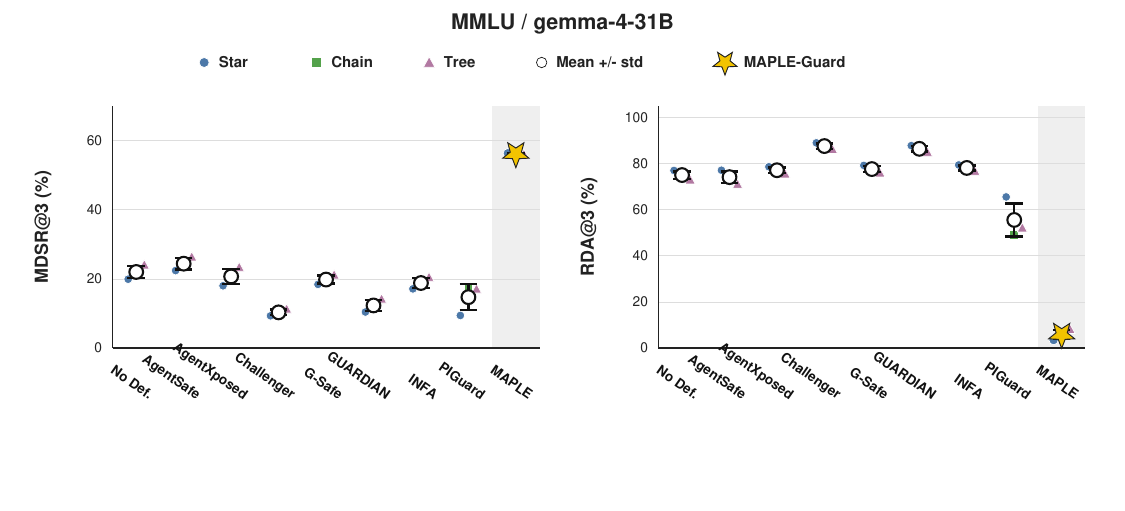}
\caption{Per-topology MDSR@3 and RDA@3 on MMLU. The upper panel uses Qwen3.5-122B-A10B, and the lower panel uses gemma-4-31B.}
\label{fig:topology_breakdown_mmlu}
\Description{Two panels compare final-round MDSR and retrieval-damage ASR across guards and star, chain, and tree topologies on mmlu. The upper panel uses Qwen3.5-122B-A10B; the lower panel uses gemma-4-31B.}
\end{figure}

\begin{figure}[!t]
\centering
\includegraphics[width=0.82\textwidth]{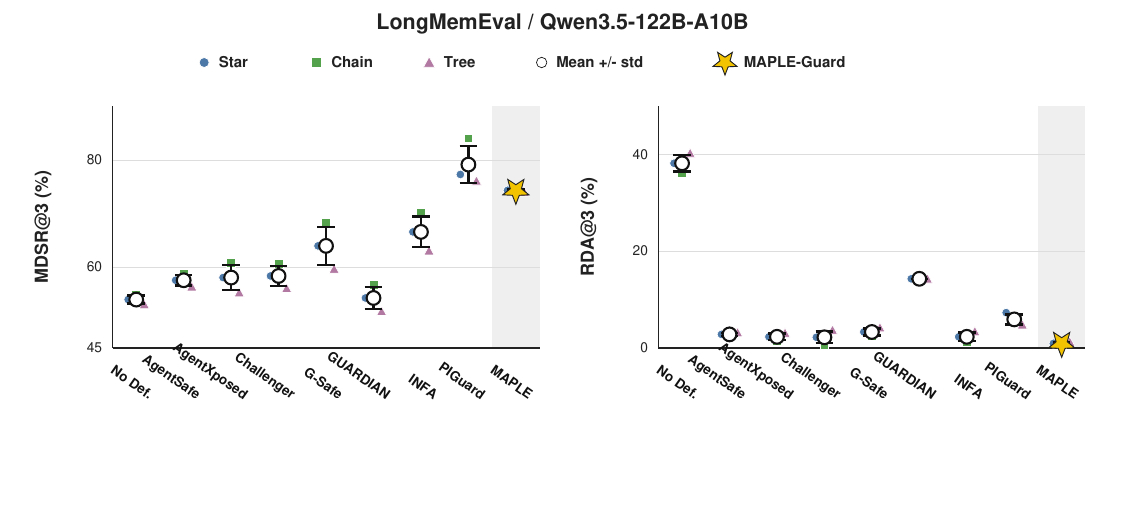}\\[-0.2em]
\includegraphics[width=0.82\textwidth]{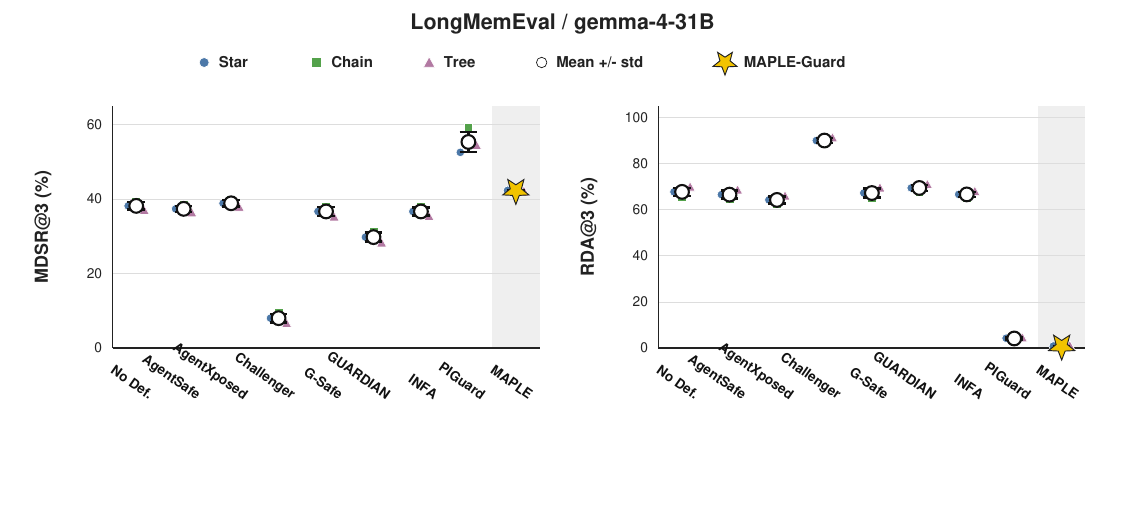}
\caption{Per-topology MDSR@3 and RDA@3 on Long\-Mem\-Eval. The upper panel uses Qwen3.5-122B-A10B, and the lower panel uses gemma-4-31B.}
\label{fig:topology_breakdown_longmemeval}
\Description{Two panels compare final-round MDSR and retrieval-damage ASR across guards and star, chain, and tree topologies on longmemeval. The upper panel uses Qwen3.5-122B-A10B; the lower panel uses gemma-4-31B.}
\end{figure}

\begin{figure}[!t]
\centering
\includegraphics[width=0.82\textwidth]{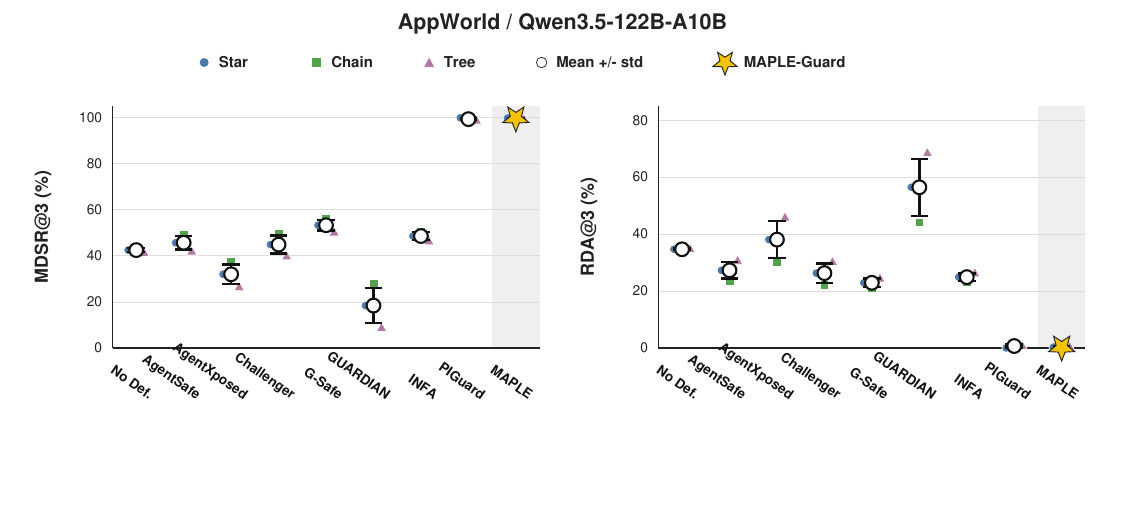}\\[-0.2em]
\includegraphics[width=0.82\textwidth]{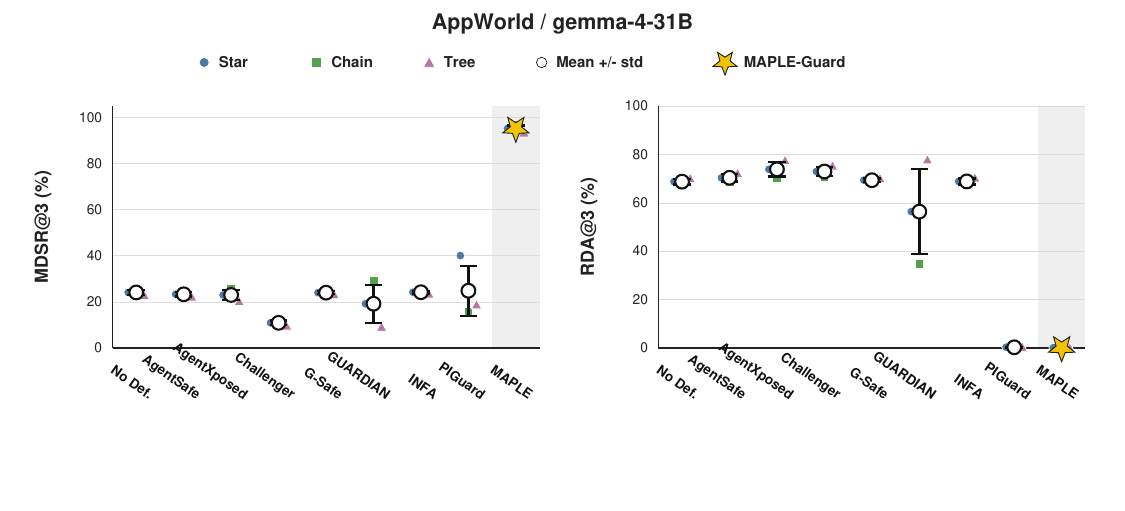}
\caption{Per-topology MDSR@3 and RDA@3 on App\-World. The upper panel uses Qwen3.5-122B-A10B, and the lower panel uses gemma-4-31B. For the upper panel, out-of-range topology markers and whisker endpoints are clipped to $[0,100]$ for display; the reported mean and standard deviation are retained.}
\label{fig:topology_breakdown_appworld}
\Description{Two panels compare final-round MDSR and retrieval-damage ASR across guards and star, chain, and tree topologies on appworld. The upper panel uses Qwen3.5-122B-A10B; the lower panel uses gemma-4-31B.}
\end{figure}

\begin{figure}[!t]
\centering
\includegraphics[width=0.82\textwidth]{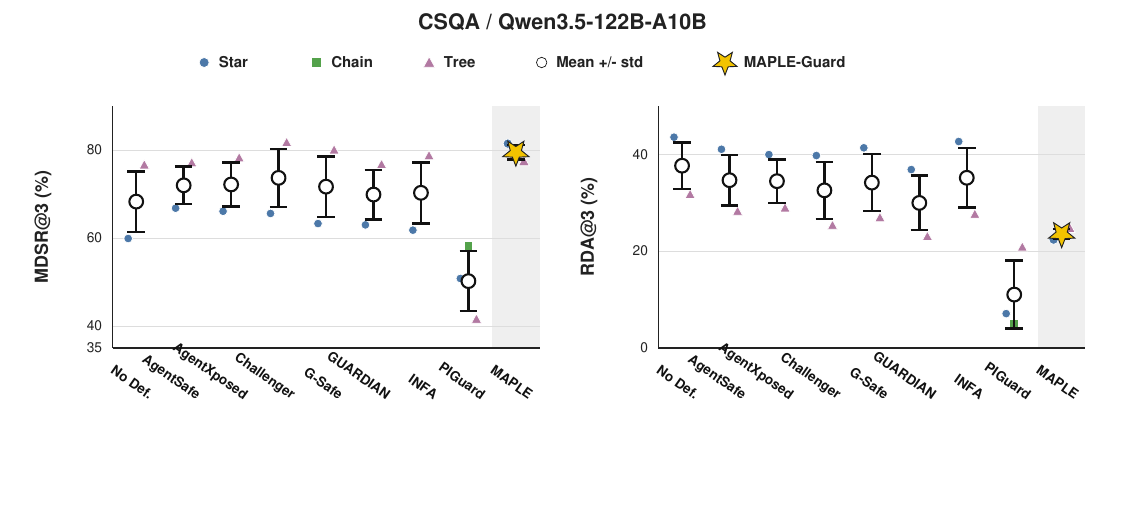}\\[-0.2em]
\includegraphics[width=0.82\textwidth]{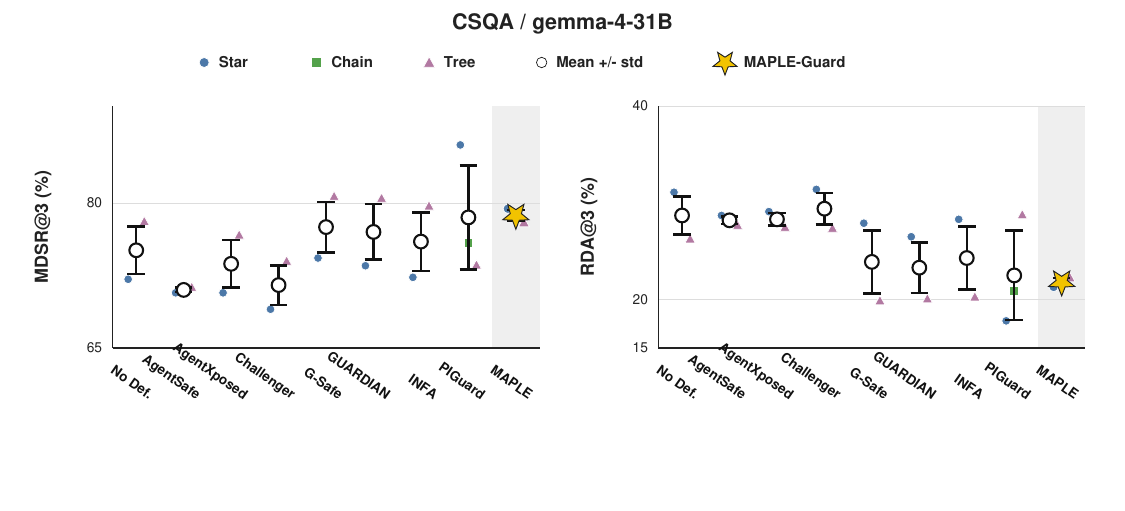}
\caption{Per-topology MDSR@3 and RDA@3 on CSQA. The upper panel uses Qwen3.5-122B-A10B, and the lower panel uses gemma-4-31B.}
\label{fig:topology_breakdown_csqa}
\Description{Two panels compare final-round MDSR and retrieval-damage ASR across guards and star, chain, and tree topologies on csqa. The upper panel uses Qwen3.5-122B-A10B; the lower panel uses gemma-4-31B.}
\end{figure}

\begin{figure}[!t]
\centering
\includegraphics[width=0.82\textwidth]{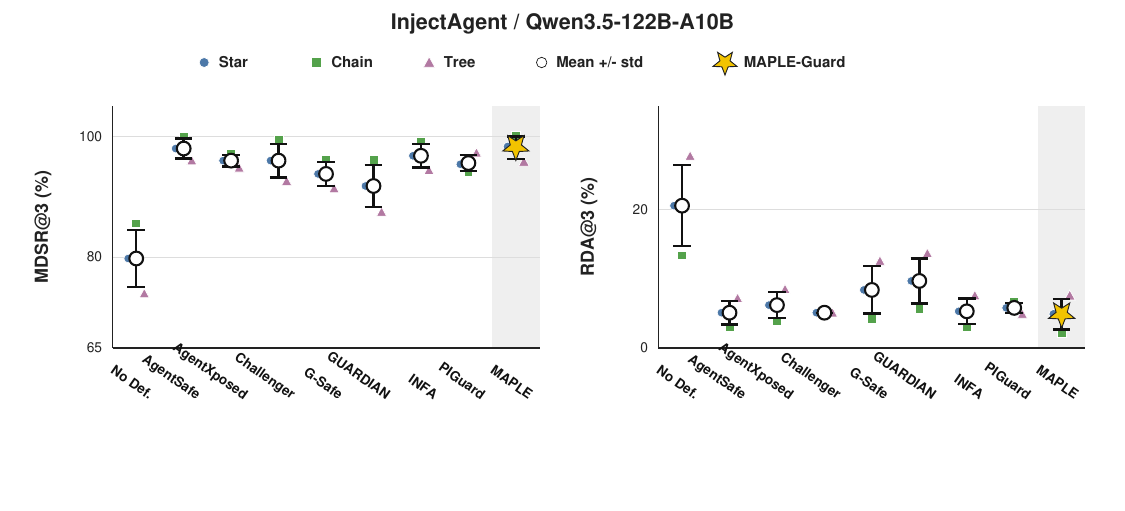}\\[-0.2em]
\includegraphics[width=0.82\textwidth]{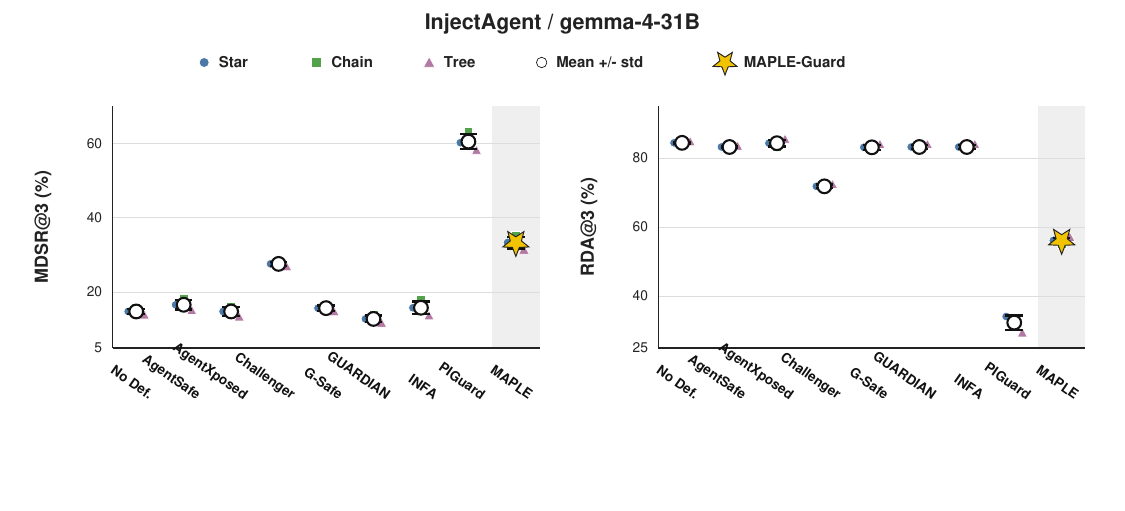}
\caption{Per-topology MDSR@3 and RDA@3 on Inject\-Agent. The upper panel uses Qwen3.5-122B-A10B, and the lower panel uses gemma-4-31B. For the upper panel, out-of-range topology markers and whisker endpoints are clipped to $[0,100]$ for display; the reported mean and standard deviation are retained.}
\label{fig:topology_breakdown_injectagent}
\Description{Two panels compare final-round MDSR and retrieval-damage ASR across guards and star, chain, and tree topologies on injectagent. The upper panel uses Qwen3.5-122B-A10B; the lower panel uses gemma-4-31B.}
\end{figure}

\FloatBarrier

\begin{table}[!t]
\caption{Round-level MMLU-MINJA results under random communication topology.}
\label{tab:mmlu_minja_round_appendix}
\centering
\scriptsize
\begin{tabular}{lcccccc}
\toprule
\textbf{Guard} & \multicolumn{6}{c}{\textbf{MMLU-MINJA}} \\
\cmidrule(lr){2-7}
& RDA@1$\downarrow$ & RDA@2$\downarrow$ & RDA@3$\downarrow$
& MDSR@1$\uparrow$ & MDSR@2$\uparrow$ & MDSR@3$\uparrow$ \\
\midrule
\multicolumn{7}{c}{\textbf{Qwen3.5-122B-A10B}} \\
\addlinespace
No Defense   & 8.6 & 8.3 & 9.7 & 86.2 & 85.2 & 82.7 \\
AgentSafe    & 7.9 & 7.8 & 7.4 & 85.6 & 84.1 & 83.6 \\
AgentXposed  & 5.3 & 5.1 & 5.6 & 77.3 & 78.7 & 78.7 \\
Challenger   & 12.6 & 11.3 & 11.0 & 80.4 & 83.1 & 82.5 \\
G-Safeguard  & 8.1 & 8.7 & 8.5 & 85.6 & 85.1 & 85.1 \\
GUARDIAN     & 8.1 & 8.0 & 8.7 & 85.1 & 85.6 & 84.6 \\
INFA-Guard   & 8.0 & 8.2 & 8.2 & 85.6 & 84.6 & 84.6 \\
PIGuard & 19.5 & 21.7 & 22.5 & 75.1 & 73.1 & 72.4 \\
\rowcolor{mapleguardgray}
\textbf{\methodname} & \textbf{0.9} & \textbf{0.4} & \textbf{1.3}
& \textbf{84.2} & \textbf{86.7} & \textbf{88.7} \\
\addlinespace
\multicolumn{7}{c}{\textbf{gemma-4-31B}} \\
\addlinespace
No Defense   & 45.0 & 58.0 & 71.2 & 54.0 & 36.5 & 20.2 \\
AgentSafe    & 45.0 & 58.2 & 71.3 & 52.5 & 35.0 & 18.5 \\
AgentXposed  & 38.5 & 49.2 & 60.0 & 51.0 & 36.0 & 21.0 \\
Challenger   & 42.0 & 53.5 & 64.3 & 53.0 & 38.0 & 23.1 \\
G-Safeguard  & 45.2 & 58.3 & 71.3 & 53.5 & 36.5 & 20.4 \\
GUARDIAN     & 45.5 & 58.6 & 71.3 & 52.0 & 35.5 & 19.5 \\
INFA-Guard   & 39.5 & 50.5 & 61.3 & 55.0 & 40.0 & 24.4 \\
PIGuard & 55.1 & 56.7 & 57.4 & 16.9 & 15.8 & 15.8 \\
\rowcolor{mapleguardgray}
\textbf{\methodname} & \textbf{2.4} & \textbf{1.7} & \textbf{1.1}
& \textbf{48.0} & \textbf{50.0} & \textbf{50.9} \\
\bottomrule
\end{tabular}

\end{table}

\begin{table}[!t]
\caption{Round-level Long\-Mem\-Eval-Memory\-Graft results under random communication topology.}
\label{tab:round_longmemeval}
\centering
\scriptsize
\begin{tabular}{lcccccc}
\toprule
\textbf{Guard} & \multicolumn{6}{c}{\textbf{Long\-Mem\-Eval-Memory\-Graft}} \\
\cmidrule(lr){2-7}
& RDA@1$\downarrow$ & RDA@2$\downarrow$ & RDA@3$\downarrow$
& MDSR@1$\uparrow$ & MDSR@2$\uparrow$ & MDSR@3$\uparrow$ \\
\midrule
\multicolumn{7}{c}{\textbf{Qwen3.5-122B-A10B}} \\
\addlinespace
No Defense   & 30.0 & 35.0 & 39.6 & 75.0 & 65.0 & 65.8 \\
AgentSafe    & 34.0 & 40.0 & 46.3 & 74.0 & 63.0 & 57.4 \\
AgentXposed  & 29.0 & 34.0 & 38.5 & 73.1 & 71.0 & 70.1 \\
Challenger   & 33.0 & 38.0 & 43.0 & 76.0 & 69.0 & 65.4 \\
G-Safeguard  & 30.0 & 35.0 & 39.1 & 78.0 & 67.0 & 66.4 \\
GUARDIAN     & 29.0 & 34.0 & 38.2 & 77.0 & 69.0 & 67.3 \\
INFA-Guard   & 31.0 & 36.0 & 40.0 & 76.0 & 64.0 & 62.7 \\
PIGuard & 6.9 & 7.1 & 7.5 & 80.2 & 79.1 & 79.1 \\
\rowcolor{mapleguardgray}
\textbf{\methodname} & \textbf{1.4} & \textbf{1.1} & \textbf{0.9}
& \textbf{78.0} & \textbf{75.0} & \textbf{73.6} \\
\addlinespace
\multicolumn{7}{c}{\textbf{gemma-4-31B}} \\
\addlinespace
No Defense   & 54.0 & 62.0 & 69.6 & 42.0 & 32.0 & 25.6 \\
AgentSafe    & 56.0 & 64.0 & 72.0 & 40.0 & 31.0 & 24.8 \\
AgentXposed  & 53.0 & 61.0 & 68.7 & 41.0 & 32.0 & 24.1 \\
Challenger   & 66.0 & 75.0 & 83.1 & 28.0 & 19.0 & 12.7 \\
G-Safeguard  & 55.0 & 63.0 & 70.4 & 41.0 & 32.0 & 24.8 \\
GUARDIAN     & 56.0 & 65.0 & 72.6 & 40.0 & 31.0 & 24.2 \\
INFA-Guard   & 52.0 & 60.0 & 67.7 & 45.0 & 36.0 & 29.0 \\
PIGuard & 6.7 & 6.7 & 7.0 & 55.7 & 55.1 & 55.0 \\
\rowcolor{mapleguardgray}
\textbf{\methodname} & \textbf{1.4} & \textbf{1.1} & \textbf{0.9}
& \textbf{38.0} & \textbf{36.5} & \textbf{35.6} \\
\bottomrule
\end{tabular}

\end{table}

\begin{table}[!t]
\caption{Round-level App\-World-Agent\-Poison results under random communication topology.}
\label{tab:round_appworld}
\centering
\scriptsize
\begin{tabular}{lcccccc}
\toprule
\textbf{Guard} & \multicolumn{6}{c}{\textbf{App\-World-Agent\-Poison}} \\
\cmidrule(lr){2-7}
& RDA@1$\downarrow$ & RDA@2$\downarrow$ & RDA@3$\downarrow$
& MDSR@1$\uparrow$ & MDSR@2$\uparrow$ & MDSR@3$\uparrow$ \\
\midrule
\multicolumn{7}{c}{\textbf{Qwen3.5-122B-A10B}} \\
\addlinespace
No Defense   & 13.0 & 16.0 & 18.7 & 93.6 & 72.5 & 74.1 \\
AgentSafe    & 9.0 & 11.0 & 12.4 & 89.8 & 76.9 & 78.5 \\
AgentXposed  & 0.0 & 0.0 & 0.0 & 98.2 & 98.2 & 98.2 \\
Challenger   & 11.0 & 13.0 & 14.5 & 89.9 & 78.8 & 75.3 \\
G-Safeguard  & 14.0 & 16.0 & 18.6 & 82.7 & 73.7 & 73.2 \\
GUARDIAN     & 15.0 & 18.0 & 20.1 & 78.5 & 74.0 & 72.2 \\
INFA-Guard   & 15.0 & 18.0 & 20.2 & 90.5 & 80.5 & 72.0 \\
PIGuard & 0.5 & 0.6 & 0.6 & 98.9 & 98.9 & 98.9 \\
\rowcolor{mapleguardgray}
\textbf{\methodname} & \textbf{0.0} & \textbf{0.0} & \textbf{0.0}
& \textbf{98.9} & \textbf{98.9} & \textbf{98.9} \\
\addlinespace
\multicolumn{7}{c}{\textbf{gemma-4-31B}} \\
\addlinespace
No Defense   & 30.0 & 35.0 & 40.2 & 68.0 & 62.0 & 57.5 \\
AgentSafe    & 31.0 & 36.0 & 40.8 & 65.0 & 59.0 & 54.0 \\
AgentXposed  & 29.0 & 34.0 & 38.1 & 67.0 & 61.0 & 56.0 \\
Challenger   & 45.0 & 51.0 & 55.9 & 32.0 & 23.0 & 17.0 \\
G-Safeguard  & 28.0 & 33.0 & 37.2 & 69.0 & 64.0 & 59.0 \\
GUARDIAN     & 30.0 & 35.0 & 39.9 & 67.0 & 62.0 & 56.5 \\
INFA-Guard   & 30.0 & 35.0 & 39.3 & 69.0 & 63.0 & 58.5 \\
PIGuard & 0.1 & 0.1 & 0.1 & 25.0 & 25.0 & 25.0 \\
\rowcolor{mapleguardgray}
\textbf{\methodname} & \textbf{2.0} & \textbf{1.5} & \textbf{1.3}
& \textbf{92.0} & \textbf{93.0} & \textbf{93.0} \\
\bottomrule
\end{tabular}

\end{table}

\begin{table}[!t]
\caption{Round-level Inject\-Agent-ToolAttack results under random communication topology.}
\label{tab:round_injectagent}
\centering
\scriptsize
\begin{tabular}{lcccccc}
\toprule
\textbf{Guard} & \multicolumn{6}{c}{\textbf{Inject\-Agent-ToolAttack}} \\
\cmidrule(lr){2-7}
& RDA@1$\downarrow$ & RDA@2$\downarrow$ & RDA@3$\downarrow$
& MDSR@1$\uparrow$ & MDSR@2$\uparrow$ & MDSR@3$\uparrow$ \\
\midrule
\multicolumn{7}{c}{\textbf{Qwen3.5-122B-A10B}} \\
\addlinespace
No Defense   & 18.0 & 21.0 & 23.7 & 91.5 & 89.0 & 87.5 \\
AgentSafe    & 25.0 & 39.0 & 51.5 & 80.0 & 60.0 & 41.5 \\
AgentXposed  & 24.0 & 37.0 & 48.1 & 78.0 & 63.0 & 54.0 \\
Challenger   & 20.0 & 24.0 & 27.3 & 91.0 & 89.0 & 88.5 \\
G-Safeguard  & 14.0 & 16.0 & 18.6 & 90.0 & 86.0 & 83.5 \\
GUARDIAN     & 13.0 & 15.0 & 16.9 & 91.0 & 90.0 & 89.0 \\
INFA-Guard   & 9.0 & 10.0 & 11.0 & 96.0 & 95.0 & 95.0 \\
PIGuard & 6.9 & 6.1 & 5.1 & 98.1 & 97.9 & 97.6 \\
\rowcolor{mapleguardgray}
\textbf{\methodname} & \textbf{5.5} & \textbf{5.3} & \textbf{5.1}
& \textbf{97.5} & \textbf{97.5} & \textbf{97.5} \\
\addlinespace
\multicolumn{7}{c}{\textbf{gemma-4-31B}} \\
\addlinespace
No Defense   & 80.0 & 84.0 & 87.0 & 18.0 & 17.0 & 16.0 \\
AgentSafe    & 80.0 & 84.0 & 87.0 & 19.0 & 18.0 & 17.0 \\
AgentXposed  & 81.0 & 85.0 & 88.0 & 18.0 & 17.0 & 16.0 \\
Challenger   & 68.0 & 73.0 & 77.0 & 34.0 & 31.0 & 28.0 \\
G-Safeguard  & 79.0 & 83.0 & 86.0 & 19.0 & 18.0 & 17.0 \\
GUARDIAN     & 80.0 & 84.0 & 87.0 & 17.0 & 15.0 & 14.0 \\
INFA-Guard   & 76.0 & 80.0 & 84.0 & 22.0 & 20.0 & 18.0 \\
PIGuard & 59.1 & 60.1 & 60.4 & 31.1 & 30.6 & 30.2 \\
\rowcolor{mapleguardgray}
\textbf{\methodname} & \textbf{54.0} & \textbf{59.0} & \textbf{62.0}
& \textbf{38.0} & \textbf{35.0} & \textbf{32.0} \\
\bottomrule
\end{tabular}

\end{table}

\FloatBarrier
\begin{table}[!t]
\caption{PIGuard round-level CSQA-Prompt\-Inject results under random communication topology. Entries are percentages.}
\label{tab:round_csqa_piguard}
\centering
\scriptsize
\begin{tabular}{lcccccc}
\toprule
\textbf{Backbone} & RDA@1$\downarrow$ & RDA@2$\downarrow$ & RDA@3$\downarrow$ & MDSR@1$\uparrow$ & MDSR@2$\uparrow$ & MDSR@3$\uparrow$ \\
\midrule
Qwen3.5-122B-A10B & 6.9 & 7.1 & 7.3 & 78.9 & 78.1 & 77.4 \\
gemma-4-31B & 20.1 & 20.4 & 20.9 & 77.1 & 76.8 & 76.5 \\
\bottomrule
\end{tabular}
\end{table}

\begin{table}[!t]
\centering
\small
\caption{PIGuard combined with full MAPLE on Gemma Base. Entries are MDSR@3 and RDA@3 (\%), averaged equally over star, chain, and tree. Standalone rows reproduce the corresponding means in main Table~\ref{tab:main_results}. The combination retains PIGuard retrieval filtering and adds MAPLE's scoring, state updates, and four lifecycle gates. Bold marks the best mean in each column.}
\label{tab:piguard_composition}
\begin{tabular*}{\textwidth}{@{\extracolsep{\fill}}lrrrr@{}}
\toprule
& \multicolumn{2}{c}{\textbf{CSQA--PromptInject}} & \multicolumn{2}{c}{\textbf{LongMemEval--MemoryGraft}} \\
\cmidrule(lr){2-3}\cmidrule(lr){4-5}
Method & MDSR@3$\uparrow$ & RDA@3$\downarrow$ & MDSR@3$\uparrow$ & RDA@3$\downarrow$ \\
\midrule
PIGuard (retrieval-only) & 78.5 & 22.5 & 55.4 & 4.1 \\
MAPLE-Guard (full) & 78.7 & 21.8 & 42.3 & \textbf{0.7} \\
PIGuard + full MAPLE & \textbf{83.3} & \textbf{18.3} & \textbf{69.6} & 1.7 \\
\bottomrule
\end{tabular*}
\end{table}
\par
\endgroup

\end{document}